\documentclass[10pt,twocolumn,letterpaper]{article}

\usepackage[pagenumbers]{cvpr} 

\usepackage{graphicx}
\usepackage{amsmath}
\usepackage{amssymb}
\usepackage{booktabs}
\usepackage{caption}
\usepackage{subcaption}

\usepackage[pagebackref=true,breaklinks=true,colorlinks,bookmarks=false]{hyperref}

\usepackage[capitalize]{cleveref}
\crefname{section}{Sec.}{Secs.}
\Crefname{section}{Section}{Sections}
\Crefname{table}{Table}{Tables}
\crefname{table}{Tab.}{Tabs.}

\begin{document}

\title{DeepFGS: Fine-Grained Scalable Coding for Learned Image Compression}

\author{Yi Ma\footnotemark[1], Yongqi Zhai\footnotemark[1], and Ronggang Wang\\
Shenzhen Graduate School, Peking University\\
{\tt\small \{mayi@pku,zhaiyongqi@stu.pku,rgwang@pkusz\}.edu.cn}
}
\renewcommand{\thefootnote}{\fnsymbol{footnote}}
\maketitle

\footnotetext[1]{These authors contribute equally.}
\begin{abstract}
Scalable coding, which can adapt to channel bandwidth variation, performs well in today's complex network environment. However, the existing scalable compression methods face two challenges: reduced compression performance and insufficient scalability. In this paper, we propose the first learned fine-grained scalable image compression model (DeepFGS) to overcome the above two shortcomings. Specifically, we introduce a feature separation backbone to divide the image information into basic and scalable features, then redistribute the features channel by channel through an information rearrangement strategy. In this way, we can generate a continuously scalable bitstream via one-pass encoding. In addition, we reuse the decoder to reduce the parameters and computational complexity of DeepFGS. Experiments demonstrate that our DeepFGS outperforms all learning-based scalable image compression models and conventional scalable image codecs in PSNR and MS-SSIM metrics. To the best of our knowledge, our DeepFGS is the first exploration of learned fine-grained scalable coding, which achieves the finest scalability compared with learning-based methods.
\end{abstract}

\section{Introduction}
Image coding technology aims to remove redundant information in images and compress image data dozens of times. With the coming of the metaverse, massive amounts of image and video data are generated and transmitted. Therefore, image and video coding technologies have become more and more important. The traditional codecs, such as JPEG \cite{1992The}, JPEG2000 \cite{2000The}, BPG \cite{bpgurl} and the latest video standard VVC \cite{VVC}, usually include four hand-crafted modules: prediction, transform coding, quantization, and entropy coding. These modules are used to remove spatial redundancy, visual redundancy and statistical redundancy in the image data, and turn the data into a stream that is easy to store and transmit.

\begin{figure}[t]
  \centering
  \includegraphics[trim={0.25cm 0.2cm 0cm 0cm},clip,scale=0.7]{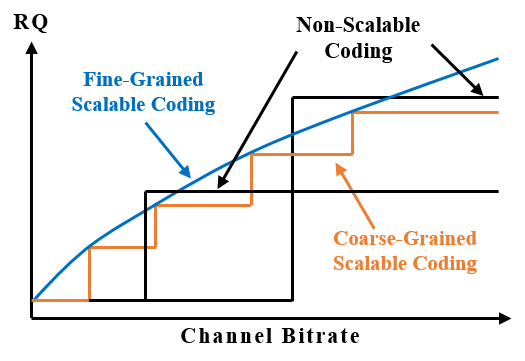}
  \caption{Illustration of coding performance for each coding technique. "RQ" indicates the image quality received by a user. Fine-grained scalable coding technique is the best solution to deal with channel bitrate changes.}
  \label{compare}
\end{figure}

\begin{figure*}
\begin{center}
\includegraphics[width=0.9\linewidth]{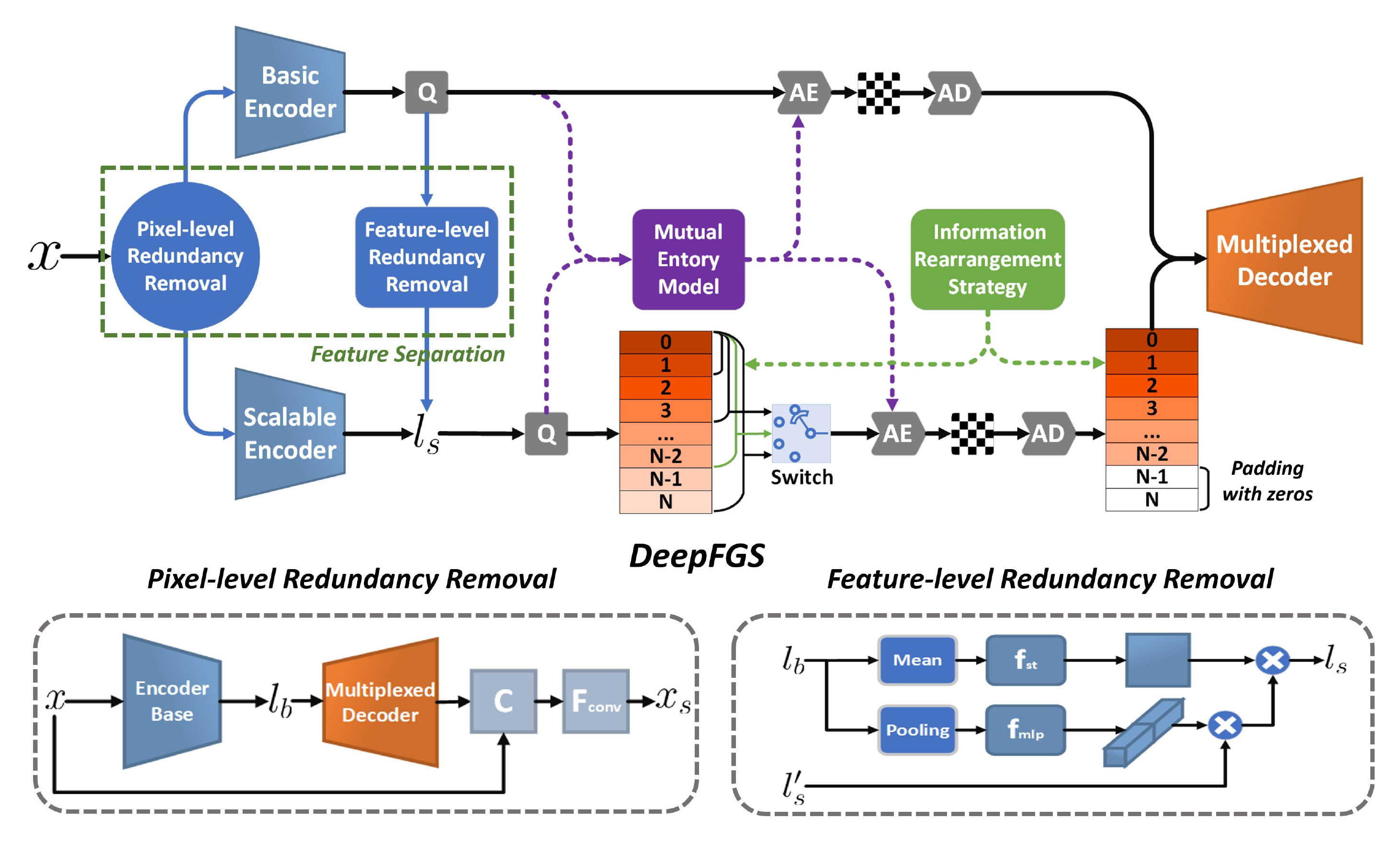}
\end{center}
  \caption{The architecture of our fine-grained scalable coding framework (DeepFGS), which includes a detailed description of the feature separation backbone.}
\label{framework}
\end{figure*}

In recent years, learning-based image compression methods have developed rapidly. Such methods rely on the image understanding and generation capabilities of the deep generation model and finally obtain better compression performance than traditional methods.
The most advanced compression model is based on a variational autoencoder (VAE), which uses an entropy model to estimate the amount of information in the compressed data, and uses a parameter to maintain the trade-off between compression efficiency and the quality of the reconstructed image.
Some existing researches \cite{balle2017end-to-end,balle2018variational,minnen2018joint,lee2019context-adaptive,cheng2020image,hu2020coarse,he2021checkerboard} have focused on designing a better entropy model to facilitate the probability estimation and get a rate-distortion performance boost. Besides, there are also studies \cite{balle2017end-to-end,liu2019non,cheng2020image} focused on improving the VAE network structure to enhance its feature extraction capabilities. 
In addition, \cite{mentzer2020high} introduces perceptual loss and adversarial training strategies to obtain reconstructed images with excellent visual quality.
Through the accumulation of the above research, learning-based image compression models have completely surpassed the traditional compression algorithms in terms of MS-SSIM, PSNR and visual quality.
Nevertheless, when we turn our focus from the pursuit of performance to actual application scenarios, image compression based on learning still lacks the realization of two important functions: variable rate and scalable bitstream.
When referring to the variable rate model, it refers to the model which can achieve different compression ratios through a single model. Traditional codecs use different quantization parameters and can easily achieve continuous variable-rate adaptation. However, the existing learned image compression methods require training multiple models to adapt to different rates. Recently, \cite{cui2021asymmetric} deployed gain units with exponential interpolation to achieve continuous rate adaptation. Although \cite{cui2021asymmetric} can achieve different compression ratios, it needs to be encoded multiple times to generate different non-scalable bitstreams. 
The role of the image compression is to convert an image into a compact bitstream, which is scalable if a subset of the bitstream can also generate a useful representation, i.e. the decoder can selectively decode part of the bitstream depending on the bandwidth.
\cite{jia2019layered} use a layered coding structure to make the bitstream coarse-grained scalable (the bitstream has four independently decodable subsets), but due to the use of multiple codecs, the inferencing process is complicated, and the scalability of the bitstream is limited. The RNN-based method \cite{toderici2015variable,toderici2017full,johnston2018improved} can generate a coarse-grained scalable bitstream through multiple iterations. However, the encoding-decoding process complexity is too high, and the rate-distortion performance is lower than the CNN-based methods.

In this paper, we propose an advanced fine-grained scalable image coding framework (DeepFGS). It can provide a highly flexible bitstream with only one encoding process to cover an entire bitrate range. This means that even if the bitstream is truncated at any position, it can be reconstructed into a complete image.
Figure \ref{compare} shows the performance difference between non-scalable coding, coarse-grained scalable coding and fine-grained scalable coding in actual bitstream transmission. It can be observed that fine-grained scalable coding can flexibly adapt to bandwidth changes and provide decoded images with the best quality.

The generation of our scalable bitstream requires the following steps: First, the image information is decomposed into basic features $l_b$ and scalable features $l_s$ through the feature separation backbone, and the redundancy between the two features is eliminated from pixel-level to feature-level. Next, the information rearrangement strategy redistributes the features in $l_s$ and establishes the forward dependence between channels instead of two-way dependence to adapt to the scalable decoding process. In this way, each additional decoded channel will bring continuous quality gain. In addition, to reduce the error in the entropy estimation process, we improve the entropy model with the mutual information of the $l_b$ and the $l_s$.
The experimental results show that our model has excellent scalability, and each subset of the bitstream can decode images of different quality. Moreover, the rate-distortion performance of our model is better than all previous scalable compression methods, whether traditional or learning-based. Our contributions can be summarized as follows:
\begin{itemize}

\item We propose the first fine-grained scalable coding framework for learned image compression that outperforms all previous scalable methods in PSNR and MS-SSIM metrics.
\item We introduce a feature separation backbone and information rearrangement strategy to obtain an extremely flexible bitstream, which can provide a continuously adjustable image quality.
\item We design a powerful mutual entropy model, which uses mutual information between the basic and scalable features to obtain more accurate probability estimation.
\end{itemize}

\section{Related Works}
\subsection{Image Compression}
Existing traditional image compression standards, such as JPEG \cite{1992The}, JPEG2000 \cite{2000The}, BPG \cite{bpgurl} and the latest VVC-Intra \cite{VVC}, typically consist of prediction, transformation, quantization, and entropy coding modules. After decades of development, most standards have been widely used in practice. However, limited by the hand-crafted design and independent optimization, it is difficult to improve the performance further.

In recent years, learning-based image compression approaches have attracted great interest with promising results. The most representative works can be divided into two branches: recurrent neural network (RNN) based models and convolutional neural network (CNN) based models. RNN based models \cite{toderici2015variable,toderici2017full,johnston2018improved} compress images or residual information from the previous step iteratively, while CNN based models typically transform images into compact latent representations for further entropy coding. Some early works \cite{theis2017lossy,balle2017end-to-end,agustsson2017soft} solve the problem of non-differential quantization and rate estimation. Afterward, some works \cite{balle2017end-to-end,balle2018variational,minnen2018joint,lee2019context-adaptive,cheng2020image,hu2020coarse,he2021checkerboard} focus on designing powerful entropy models to improve the accuracy of rate estimation. Some approaches use generalized divisive normalization (GDN) layers \cite{balle2017end-to-end}, or attention maechanism \cite{liu2019non,cheng2020image}, or adaptive feature extration models \cite{ma2021afec} to improve the variational autoencoder (VAE) architechure. Recently, several studies have been done to improve the subjective quality. \cite{rippel2017real,agustsson2019generative,mentzer2020high} use generative models and perceptual loss to generate vivid texture. \cite{mentzer2018conditional,zhang2021attention,cai2019end,ma2021variable} introduce importance map or region-of-interest (ROI) mask to allocate more bits to the important/ROI areas.

All the above methods are dedicated to achieving higher compression efficiency and better visual quality. However, in many real-time applications, bandwidth resources fluctuate drastically, most codecs cannot adapt the bitrate based on the network conditions. The best solution is to make the bitstream scalable and flexible, which means that the decoder is able to decode parts of the stream to reconstruct the image with different quality. 

\subsection{Traditional Scalable Image Compression}
The scalability of the bitstream is supported by many traditional standards, such as JPEG2000 \cite{2000The}, the H.264/AVC scalable extension (Scalable Video Coding, SVC) \cite{wiegand2007text,schwarz2007overview} and the HEVC scalable extension (SHVC) \cite{boyce2015overview}. These standards encode the image once as several dependent bitstreams, including a base layer and several enhancement layers. During transmission, the encoder first sends the bitstreams of the base layer to reconstruct the basic quality of the image, and if the bandwidth is sufficient, it sends the bitstreams of the enhancement layers in turn. The more bits are received, the better the quality of image reconstruction is. The scalability of image compression mainly includes spatial scalability, fidelity scalability, and fine granularity scalability (FGS) \cite{schwarz2007overview,boyce2015overview,li2001overview}. Spatial scalable compression generates bitstreams with different resolutions. Fidelity scalability can be regarded as a special case of spatial scalability, which generates images of different quality levels. FGS encodes the image into a base layer bitstream and an enhancement layer bitstream that can be truncated anywhere. Specifically, the enhancement layer uses bit-plane decomposition to encode the residuals of the original image DCT \cite{ahmed1974discrete} coefficients and the quantized DCT coefficients of the base layer. In bit-plane decomposition, the higher bits (important parts) of each coefficient are coded first, thus FGS can finely adjust image quality according to the number of bits received.

\subsection{Learned Scalable Image Compression}
In the early stage, \cite{toderici2015variable} uses RNN structure to compress raw and residual information iteratively. After each iteration, the bitstreams will increase, and the quality of the reconstructed image will be enhanced. Following that, \cite{su2020scalable} presents an RNNs architecture with quantization and entropy coding to achieve scalable learned image compression. Inspired by the conventional scalable coding, \cite{jia2019layered} designs a scalable auto-encoder to compress the original images and residual images iteratively. \cite{zhang2019learned} adopts bit-plane decomposition and bidirectional ConvLSTM to achieve scalable coding. \cite{guo2019deep} proposes a four layers scalable compression framework and uses a decorrelation unit to eliminate redundancy between previous layers and current layers. \cite{mei2021learning} designs a spatial/quality scalable compression framework with latent-feature reuse and prediction.

However, there are still some disadvantages in the above methods. Firstly, these methods cannot achieve fine granularity scalability. The number of layers is limited, and the bitrates gap between layers is quite large, which is too coarse for changes in channel bandwidth. Secondly, they have high computational complexity. Due to the complex iterative manner, RNN based models suffer from highly time and memory-consuming. Most methods have multiple codecs and require multiple complete encoding and decoding processes to achieve scalability, which is too complicated in practical applications.

\section{Proposed Method}
In this section, we introduce our proposed fine-grained scalable image compression framework. As shown in Figure \ref{framework}, the feature separation backbone divides the image information into basic features $l_b$ and scalable features $l_s$. Then through the information rearrangement strategy, the information in the latent representation is redistributed. We have deployed feature selection switches and feature fusion modules in the front of the decoder to implement scalable decoding and decoder reuse. The details of the network are given in the Appendix.
\subsection{Feature Separation Backbone}

In order to achieve scalable coding, only a certain subset of the bitstream can be used to reconstruct the image, which will cause the image quality to be spatially uneven. To this end, we divide the features of the image into two parts. The first is the basic features $l_b$ that contribute the indispensable basic texture of the image, and the second part is the scalable features $l_s$ that bring optional quality improvement. In this way, $l_b$ guarantees the basic quality, and $l_s$ makes the bitstream flexible. The redundancy between $l_b$ and $l_s$ will be removed from pixel domain to feature domain in two steps. First, as shown in Figure \ref{framework}, image $x$ is sent to the basic encoder $g_a$ to extract the basic feature $l_b$.
\begin{equation}
l_b = Q(g_{a}(x)) 
\label{equation1}
\end{equation}
where $Q$ denotes the quantization operation. And the initial scalable features $x_{s}$ will be obtained by the difference operation between $x$ and $\hat{x}$ (the reconstruction of $l_b$). 
\begin{align}
&   x_{s} =F_{conv}(x || \hat{x})   \qquad with \quad \hat{x} = g_s(l_b)
\label{equation2}
\end{align}
where $g_s$ and $||$ denotes the decoder and the concat operation. $F_{conv}$ refers to the convolution operation, which is used to squeeze the channel. Although the pixel-level redundancy in $x$ and $x_{s}$ has been removed, there will be feature-level redundancy when they are compressed into latents respectively. Figure \ref{framework} shows our feature-level redundancy removal module (FRR). We use $l_b$ as a cross guidance to filter the redundant information in the $l_s$ in the channel and space domain.

\begin{align}
   l_{s} =l_{s'} \otimes f_{mlp}(pooling(l_{b})) \otimes f_{st}(mean(l_{b}))&\\\nonumber
    \qquad with \quad l_{s'} =g_{a1}(x_s)&
\label{equation3}
\end{align}
where $g_{a1}$ denotes the scalable encoder in Figure \ref{framework}. $pooling$ and $mean$ refer to the average in channel and space, respectively. $\otimes$ denotes the channel-wise multiplication, $f_{mlp}$ denotes multi-layer perceptron, and $f_{st}$ denotes the operation that first unsqueezes the channel and then squeezes it to 1.
In this way, we have obtained compact latent representations $l_s$ and $l_b$, and then we will redistribute the features in $l_s$ through the information rearrangement strategy.

\begin{figure}
  \centering
  \includegraphics[scale=0.12]{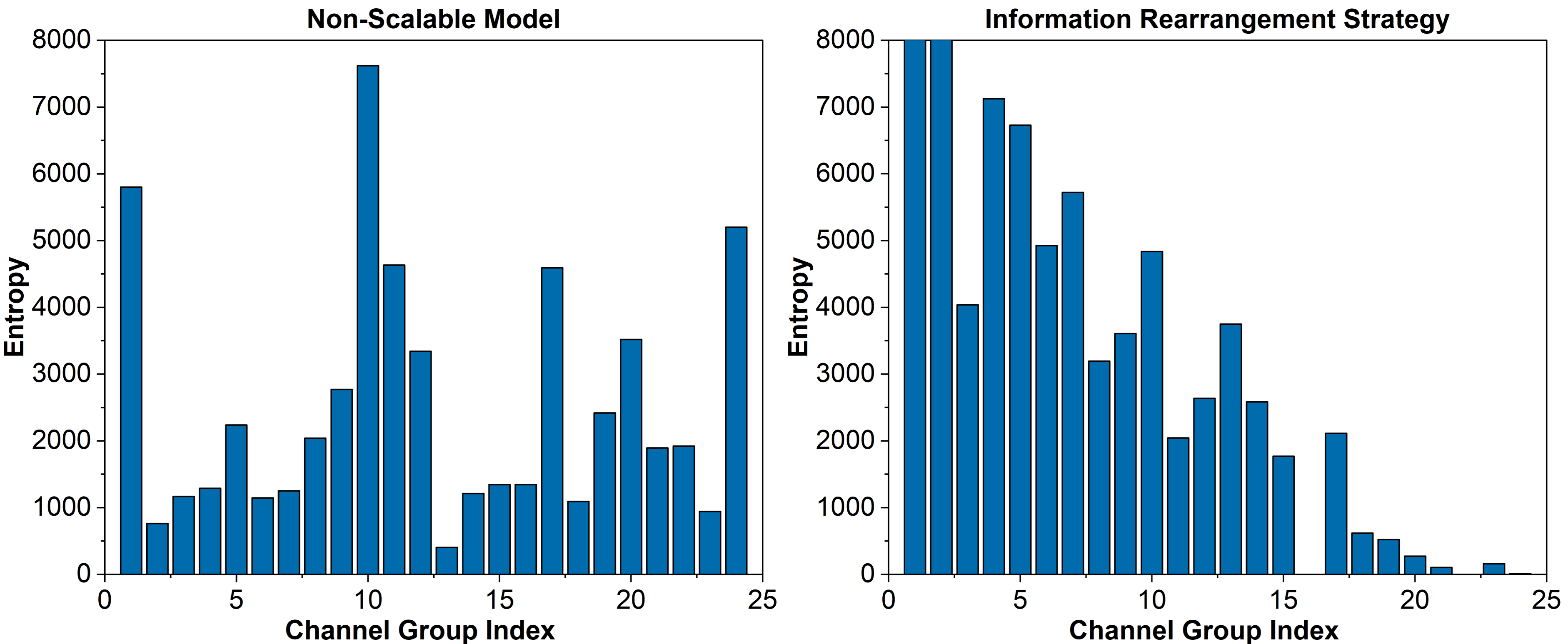} 
  \caption{Visualization of entropy of the latent representation from kodim20. We divide the 192 channels into 24 groups on average, and the figure shows the entropy of each group.}
  \label{entropy}
\end{figure}

\subsection{Information Rearrangement Strategy}

In the learning-based image compression framework, the features of the image $x$ are extracted into the latent representation $l \in \mathbb{R}^{C \times H \times W}$ where $c$, $h$, and $w$ represent the number of channels, height and width of the feature map. As shown in Figure \ref{entropy} the contribution of $l$ to the reconstruction quality is not channel-wise uniform. We use the information rearrangement strategy to redistribute the features in $l_s$ and establish the forward dependence between channels instead of two-way dependence, which means that each channel only needs to be combined with the previous channel to complete the reconstruction. As shown in Figure \ref{2-psnr}, establishing the forward dependency allows us to obtain continuous quality improvement when reconstructing through the first $n$ channels (remaining channels are padding with zeros). Instead of explicitly rearranging features, we designed an optimization function to guide network learning how to do so.
In our framework, the features to be encoded include $l_b \in \mathbb{R}^{C_1 \times H \times W}$ and $l_s \in \mathbb{R}^{C_2 \times H \times W}$, both $C_1$ and $C_2$ is set to 192. Let $k\in [C_1,C_2]\cap \mathbb{N}$ denote the number of channels available for decoding, up to 384, which simulates bandwidth fluctuations. When $k$ is equal to 192, it means that only the basic feature $l_b$ is available. Therefore, the optimization function is:

\begin{figure}
  \centering
  \includegraphics[scale=0.14]{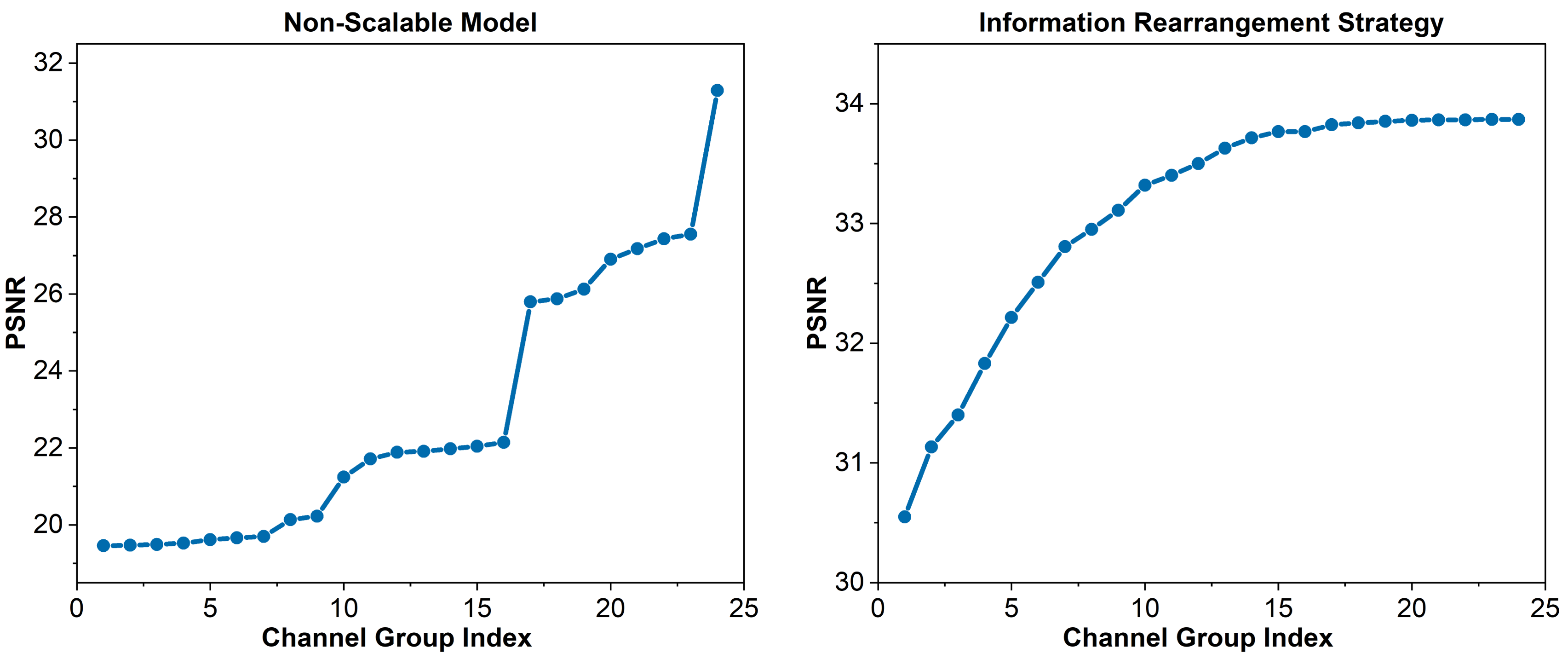} 
  \caption{Reconstruction quality of the latent representation of kodim20. We divide the 192 channels into 24 groups on average. The abscissa represents the first $n$ groups, and the ordinate represents the quality of the reconstructed image through of the first $n$ groups.}
  \label{2-psnr}
\end{figure}

\begin{equation}
J_{base} =R(l_b)+\lambda D(x,g_s(l_b))
\label{equation4}
\end{equation}
$D$ denotes distortion metric (such as MSE or MS-SSIM), $R$ refers to probability estimation(see section \ref{Entropy_Model}), used to calculate the number of bits, and $\lambda$ is used to balance them. When $k$ is greater than 192, additional scalable features can be obtained, and we need to optimize each forward-dependent channel combination:
\begin{align}
\label{equation_j}
 &  J_{scalable} =R(l_b)+\lambda D(x,g_s(l_b)) + \\\nonumber 
 &  \qquad \sum_{i=1}^{k-C_1} R(l_b) +R(l_s^i) + \lambda w(i) D(x,g_s(l_b || l_s^i))
\end{align}
Where $l_s^i$ refers to the first $i$ channels in $l_s$. $w(\cdot)$ determines the bitrate range covered by our model, which can adjust the scalability of the bitstream, here we set $w(i) = i Mod 8$. Note that the same decoder $g_s$ is used in Equation \ref{equation2} and Equation \ref{equation_j}, because the multiplexing of the decoder can reduce the complexity of the model. However, the optimization of the second term in Equation \ref{equation_j} requires dozens of forward propagation and back propagation, resulting in unacceptably high computational complexity. Therefore, we simplify it by replacing the accumulation operation with the sampling operation.

\begin{figure}[t]
  \centering
  \includegraphics[scale=0.25]{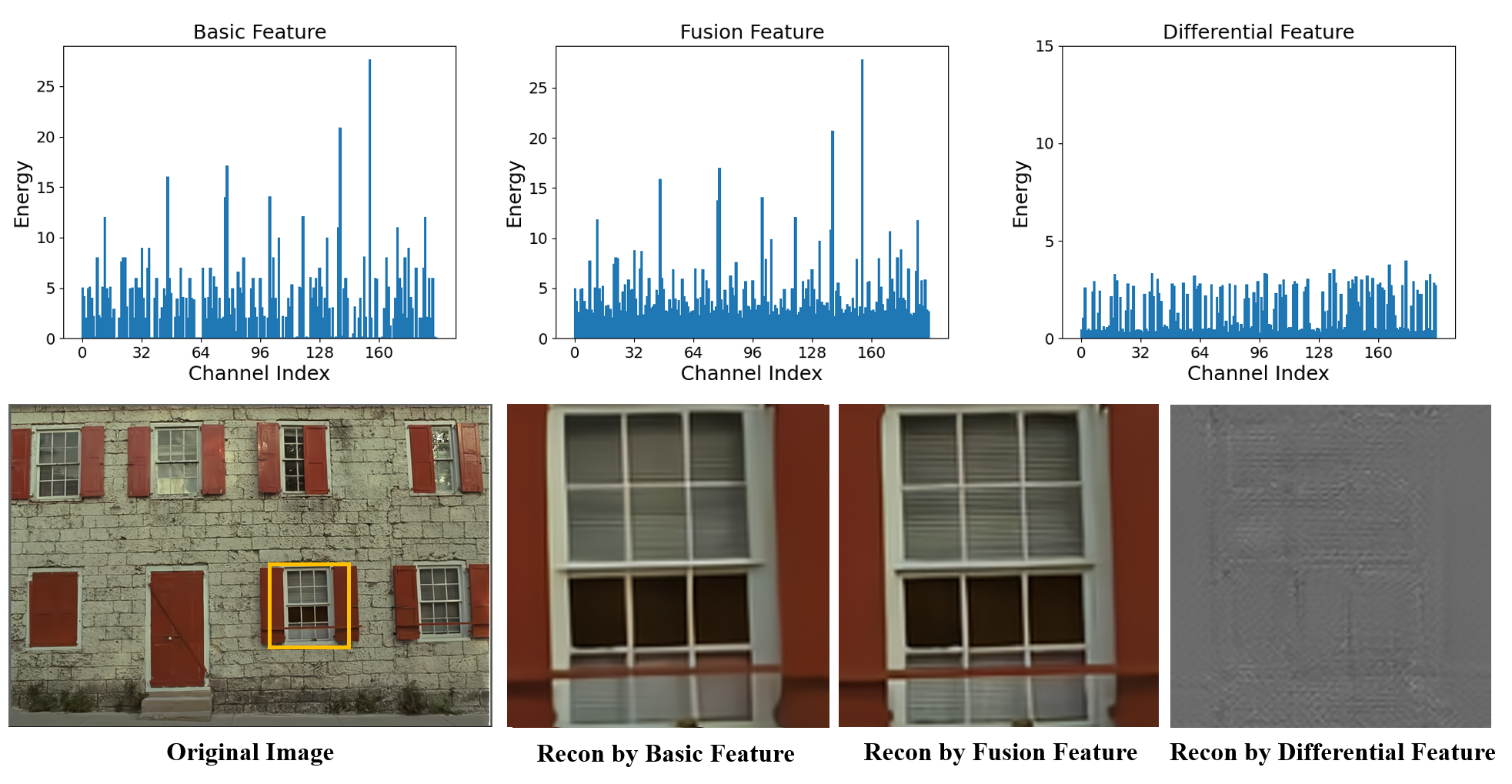} 
  \caption{Observation of the process of feature fusion. Basic feature, fusion feature, and difference feature refer to: $FFM(l_b)$, $FFM(l_b || l_s)$ and $FFM(l_b || l_s) - FFM(l_b)$. We search for the feature map with the largest activation value among the features and visualize the energy of each channel ($maximum-minimum$).}
  \label{view_feature}
\end{figure}

\begin{figure*}[t]
    \centering
  \begin{subfigure}[]{0.39\textwidth}
        \centering
      \includegraphics[width=1\textwidth]{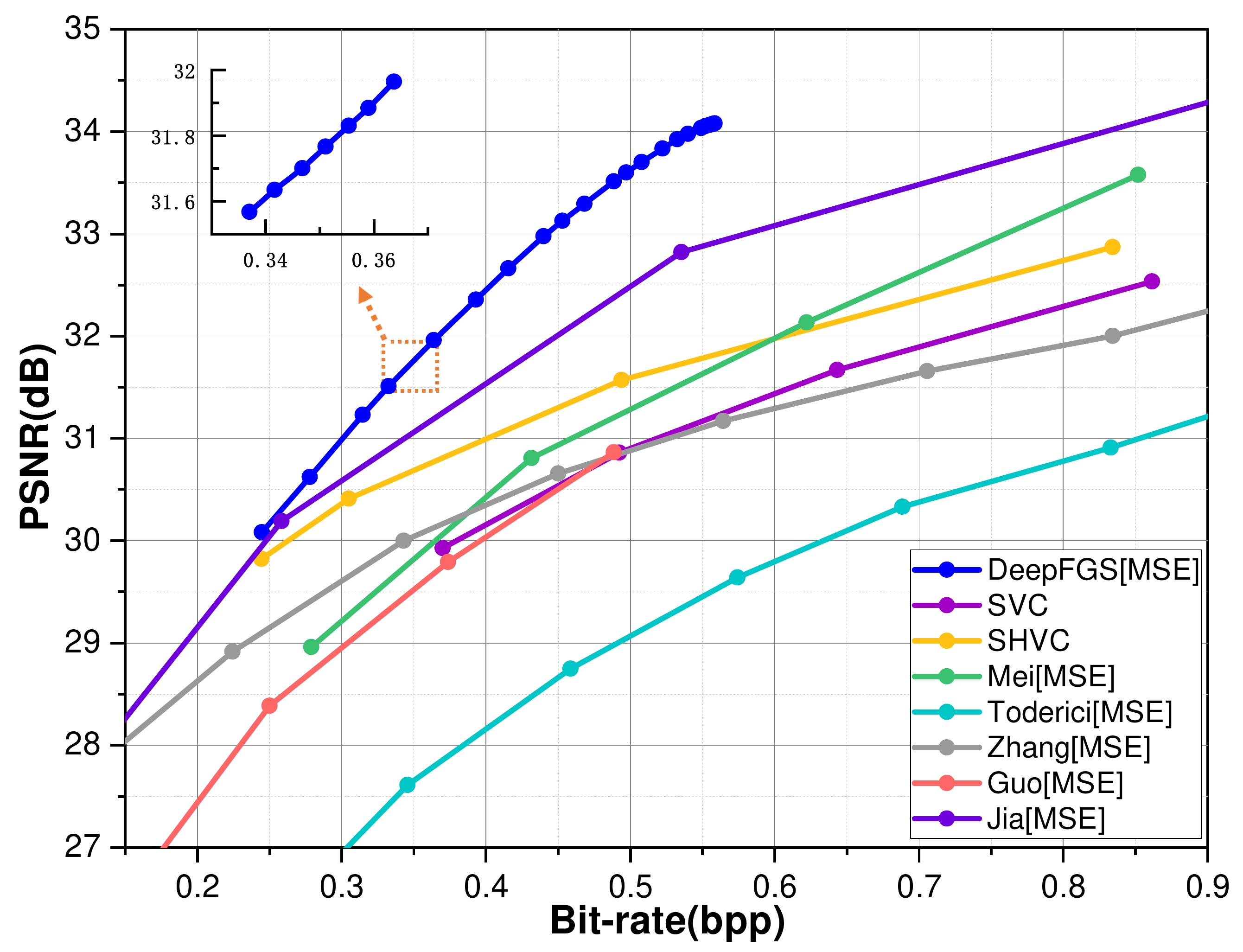} %
        \caption{}
        \label{subfigure_1}
    \end{subfigure}
    \hspace{10mm}
  \begin{subfigure}[]{0.39\textwidth}
        \centering
      \includegraphics[width=1\textwidth]{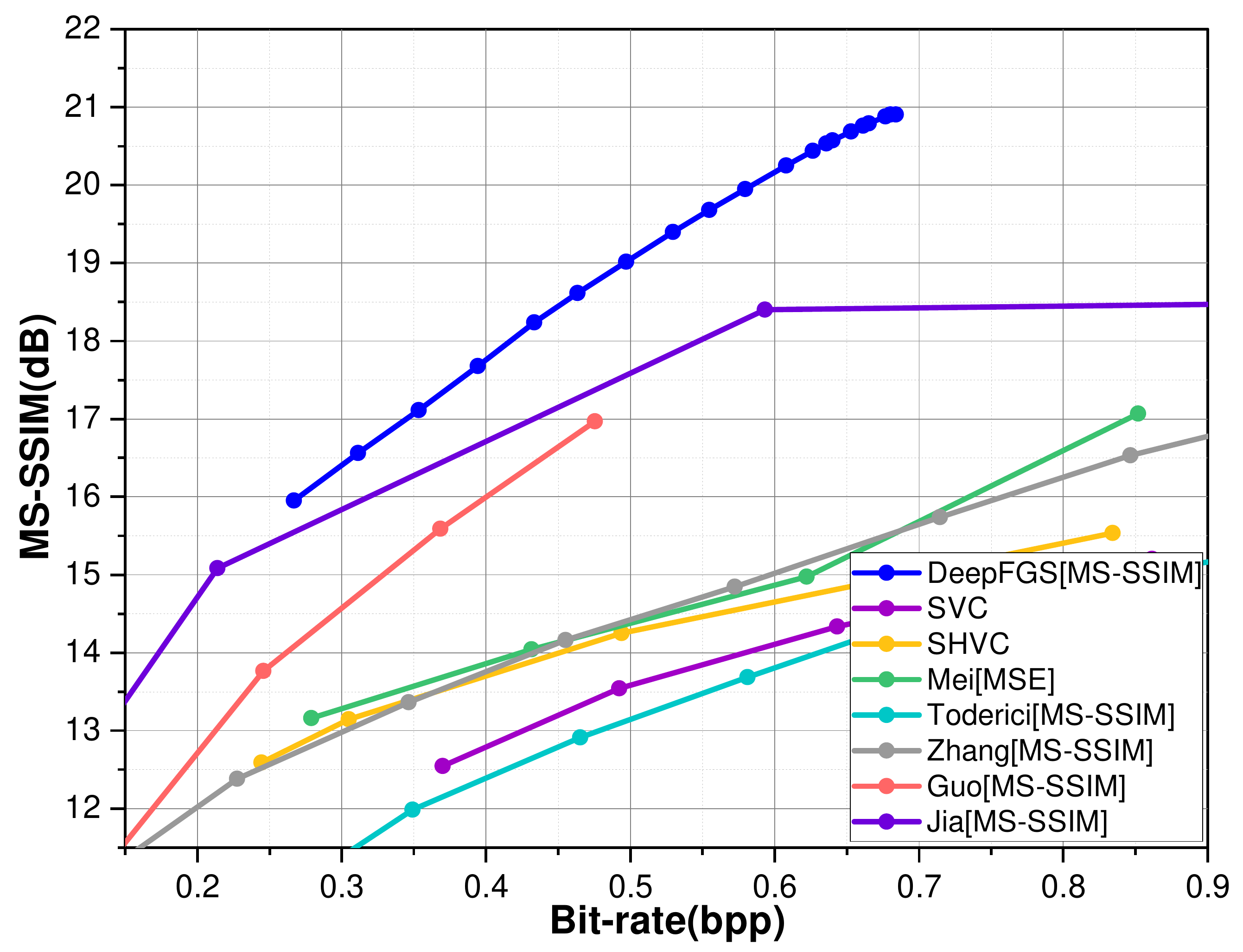}
        \caption{}
        \label{subfigure_2}
    \end{subfigure}
    \hfill
  \begin{subfigure}[]{0.39\textwidth}
        \centering
      \includegraphics[width=1\textwidth]{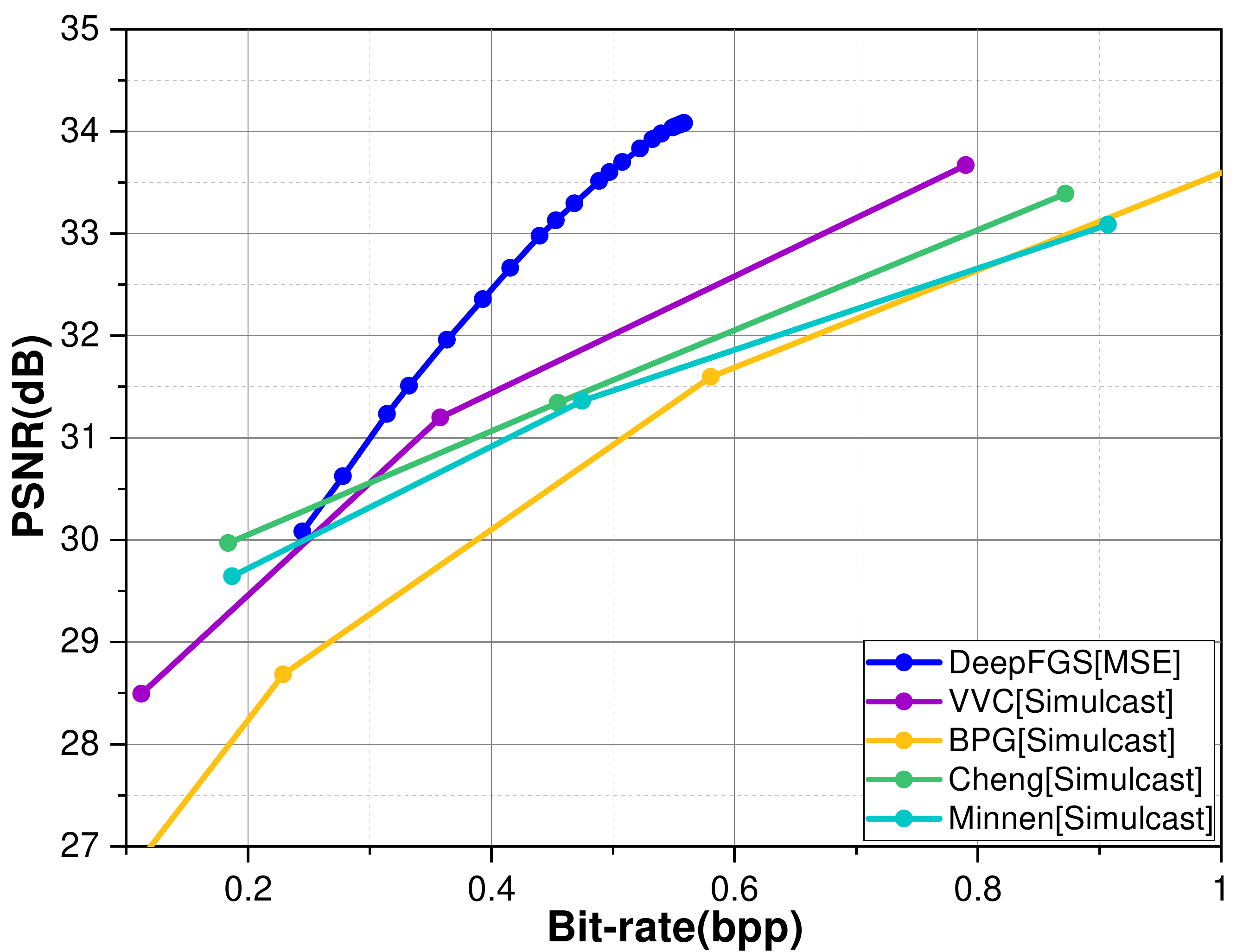}
        \caption{}
        \label{subfigure_3}
    \end{subfigure}
    \hspace{10mm}
  \begin{subfigure}[]{0.39\textwidth}
        \centering
      \includegraphics[width=1\textwidth]{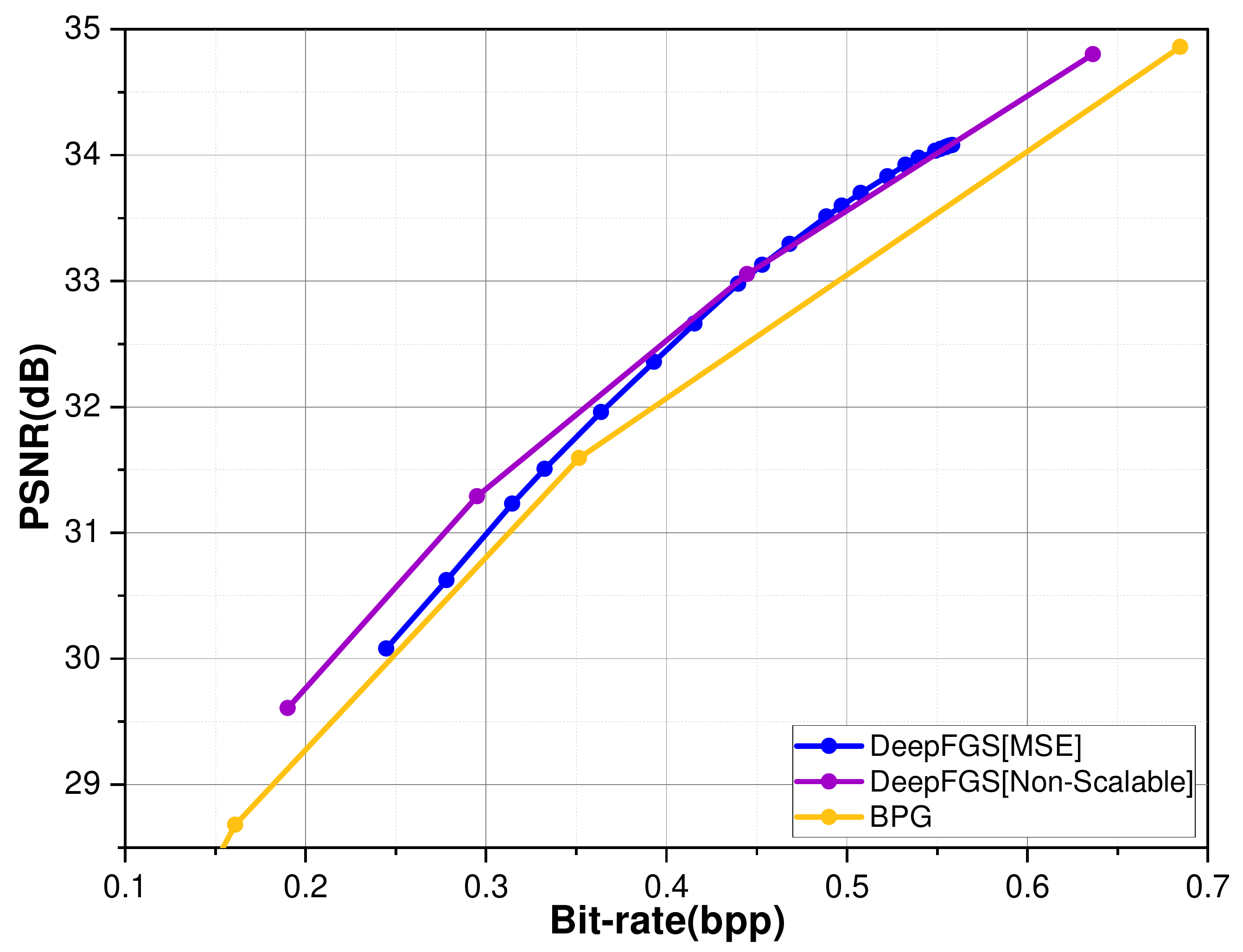}
        \caption{}
        \label{subfigure_4}
    \end{subfigure}
\caption{Comparison of rate-distortion performance of our DeepFGS with existing compression standards and other learned compression models, including Toderici \cite{toderici2015variable}, Zhang \cite{zhang2019learned}, Guo \cite{guo2019deep}, Jia \cite{jia2019layered}, Mei \cite{mei2021learning}. (a): PSNR performance evaluation on Kodak dataset: compare with scalable coding models. (b): MS-SSIM performance evaluation on Kodak dataset: compare with scalable coding models. (c): PSNR performance evaluation on Kodak dataset: compare with simulcast coding models. (d): PSNR performance evaluation on Kodak dataset: compare with non-scalable coding models.}
\label{fig:RMSE-100run}   
\end{figure*}

\begin{align}
 &  J_{scalable} =R(l_b)+\lambda D(x,g_s(l_b)) + \\\nonumber 
 &  \qquad  R(l_b) +R(l_s^j) + \lambda w(j) D(x,g_s(l_b || l_s^j))\\\nonumber 
\label{equation6}
\end{align}
where $j$ is sampled from $[C_1,C_2]\cap \mathbb{N}$.
We design two experiments to illustrate the role of information rearrangement strategies. In Figure \ref{entropy}, we show the amount of information in each channel in the latent representation. It can be observed that the information distribution of the non-scalable image compression network is messy and irregular, and our model arranges the channels in descending entropy order. Only in this way can we achieve continuous and smooth growth of quality. Figure \ref{2-psnr} clearly shows the impact of the information rearrangement strategy on the quality of the reconstructed image. The non-scalable compression model is volatile when using the first $n$ channels for decoding, while our model maintains a smooth and steady growth in an extensive bit rate range.

\subsection{Decoder Multiplexing and Analysis}
Many studies on learned scalable image compression, such as \cite{jia2019layered} and \cite{mei2021learning}, generate bitstreams of different quality levels, which use more than two decoders to reconstruct these bitstreams. Considering that these decoders all try to reconstruct the same input image and are optimized by similar loss functions, their reconstruction modes are similar. Therefore, the complexity can be reduced by multiplexing a part of the decoder parameters. As shown in Figure \ref{framework}, our model includes two decoding processes, which are located in the feature separation backbone and at the end of the entire framework. In order to reduce the parameters and complexity of the decoding network, the two locations share the same decoder. Therefore, we design a feature fusion module (FFM) at the head of the decoder.
\begin{align}
  l_{m}=l_{in} \otimes f_{mlp}(pooling(l_{s})) \otimes f_{st}(mean(l_{s}))& \\\nonumber
    \qquad with \quad l_{in} =Swich(l_b, l_b || l_s)&
\label{equation7}
\end{align}
where $l_m$ denotes the fusion feature, which is passed to the subsequent network for reconstruction. 

In addition, we design an observation experiment to demonstrate the process of feature fusion. First, from the first row of Figure \ref{view_feature}, we can observe that the basic feature and the fusion energy distribution are similar. The difference (differential feature) between the two is reflected in some low-energy channels. By comparing with the reconstructed image in the second row, we know that the high-energy channels of the basic features carry low-frequency basic textures. The differential features include high-frequency textures that can improve the quality of the reconstruction, which is consistent with our expectations for the feature separation backbone.

\begin{figure}
  \centering
  \includegraphics[scale=0.26]{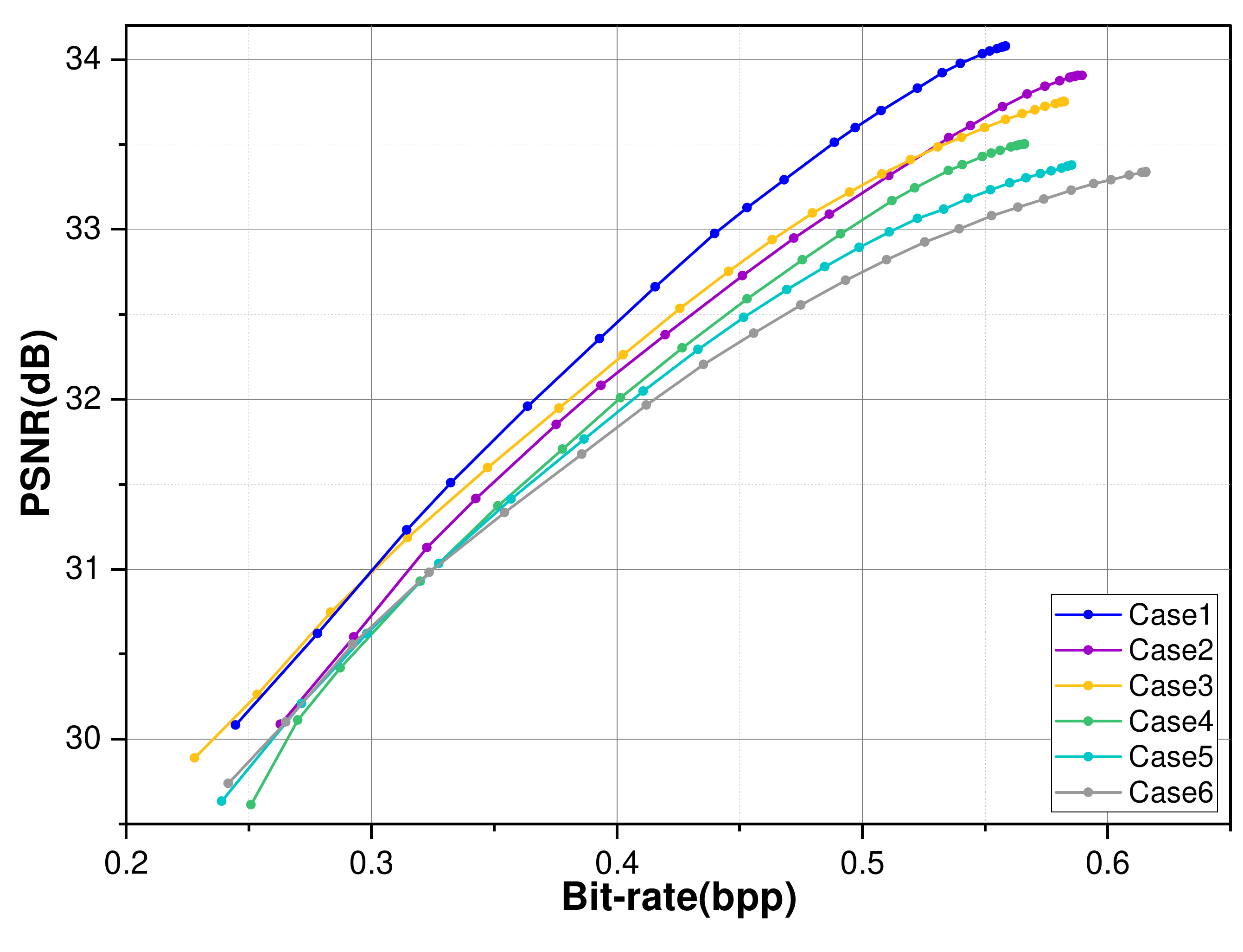}
  \caption{Ablation study on the proposed DeepFGS. The specific ablation configurations are shown in Table \ref{table1}.}
  \label{ablation_pic}
\end{figure}

\subsection{Quantization and Mutual Entropy Model}
\label{Entropy_Model}
Let ${x}$ denotes the original image, ${l}$ is the latents generated by the encoder, and after quantization $\hat{l}= Q(l)$. In order to make the model end-to-end trainable, we approximate the quantization operation by adding uniform noise \cite{balle2017end-to-end} during training. 

Following hyperpriors \cite{balle2018variational}, the rate of quantized latents is estimated by the mean and scale Gaussian entropy model, which can be formulated as:
\begin{equation}
  p_{\hat{l}|\hat{z}}(\hat{l}|\hat{z}) =  \mathcal{N}(\mu,{\sigma}^2)
  \label{eq:prob_y}
\end{equation}
where ${z}$, $\hat{z}$ represent the hyper latents before and after quantization, ${\mu}$, ${{\sigma}^2}$ are the estimated mean and scale parameters of latents $\hat{l}$. In our model, $\hat{l}_b$, $\hat{z}_b$ are the basic latents, and $\hat{l}_s$, $\hat{z}_s$ are the scalable latents. The rates of ${l}_b$ and ${l}_s$ can be rewritten as :
\begin{align}
  R_{{b}} &= \mathbb{E}[-{\log_2}p_{\hat{l}_b}({\hat{l}_b})] + \mathbb{E}[-{\log_2}p_{\hat{z}_b}({\hat{z}_b})]
  \label{eq:rate_y}\\\nonumber
  R_{{s}} &= \mathbb{E}[-{\log_2}p_{\hat{l}_s}({\hat{l}_s})] + \mathbb{E}[-{\log_2}p_{\hat{z}_s}({\hat{z}_s})] + R_{{b}} 
\end{align}

In the previous methods \cite{jia2019layered,mei2021learning}, they use independent entropy models for the base and enhancement layers. However, due to the existence of redundancy between latents, the mutual information ${I(\hat{l}_s, \hat{l}_b)}$ is positive. Let ${H(\hat{l}_s)}$ denotes the Shannon entropy of $\hat{l}_e$, then the conditional entropy ${H(\hat{l}_s|\hat{l}_b)}$can be obtained as:
\begin{align}
  H(\hat{l}_s|\hat{l}_b) &= H(\hat{l}_s) - I(\hat{l}_s, \hat{l}_b) < H(\hat{l}_s)
  \label{ih}
\end{align}

\begin{table}
\begin{center}
\caption{The specific configuration of ablation study.}
\label{table1}
\resizebox{50mm}{15mm}{
\begin{tabular}{cccc}
\toprule
 & $FRR$ & $FFM$ & $MEM$ \\ 
\midrule
\textbf{$Case 1$}             & \checkmark     & \checkmark     & \checkmark           \\
\textbf{$Case 2$}             & \checkmark    & \checkmark     &            \\
\textbf{$Case 3$}             &\checkmark     &      &\checkmark            \\
\textbf{$Case 4$}             &     & \checkmark     &            \\
\textbf{$Case 5$}             &      &      & \checkmark           \\
\textbf{$Case 6$}             &      &      &            \\

\bottomrule
\end{tabular}}
\end{center}
\vspace{-1.5em}
\end{table}

\begin{figure*}
\begin{center}
\includegraphics[width=1.0\linewidth]{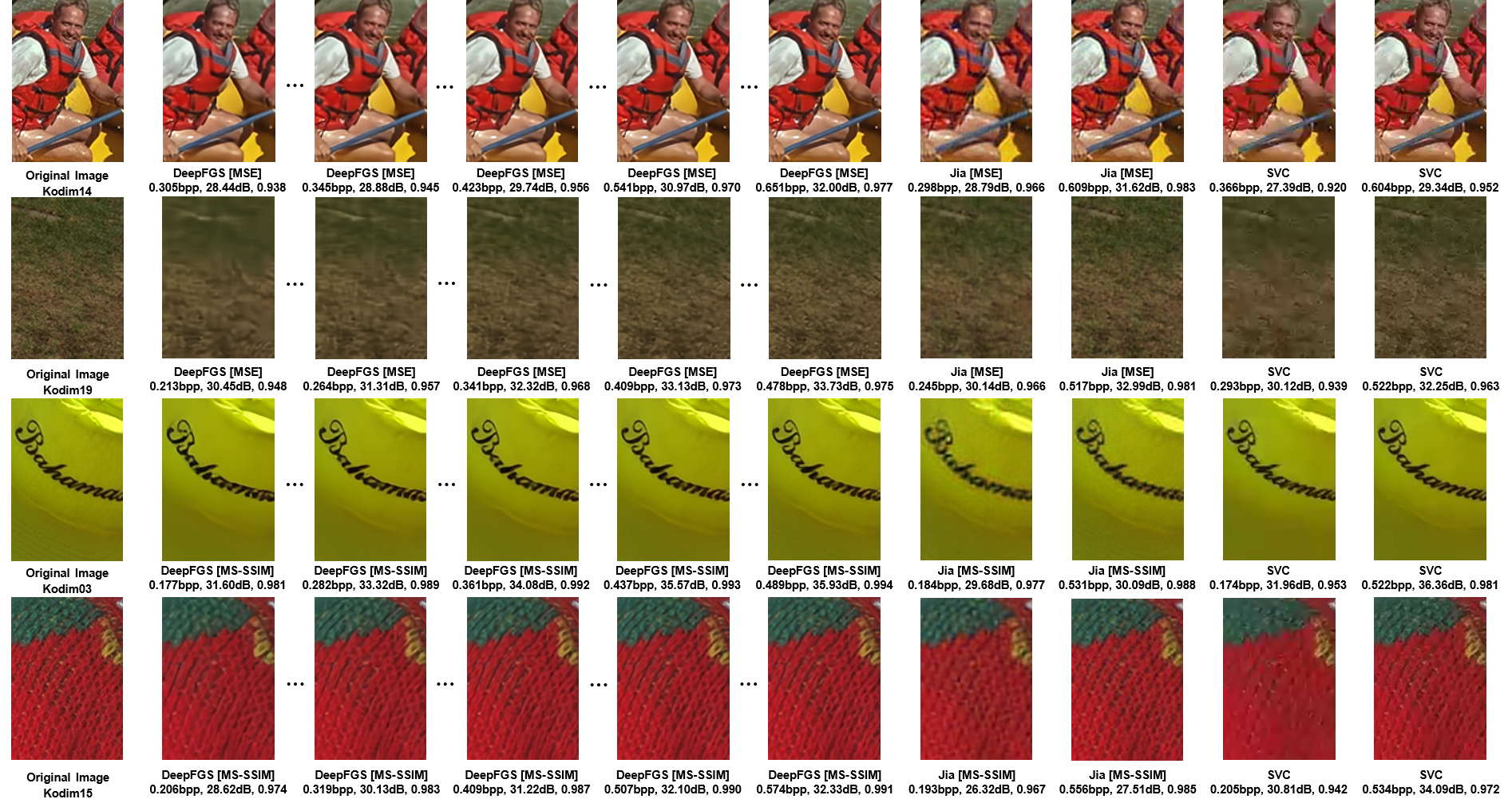}
\end{center}
  \caption{Visual quality comparisons with the learning-based layered scalable method Jia \cite{jia2019layered} and traditional scalable codec SVC.}
\label{visual_results}
\end{figure*}

Therefore, in our entropy model, $\hat{l}_s$ is not only conditioned on the information of $\hat{z}_s$, but also uses $\hat{l}_b$ as the prior information. The equation \ref{eq:rate_y} can be formulated as:
\begin{align}
  R_{{b}} = \mathbb{E}[-{\log_2}p_{\hat{l}_b|\hat{z}_b}({\hat{l}_b|\hat{z}_b})] + \mathbb{E}[-{\log_2}p_{\hat{z}_b}({\hat{z}_b})]
  \label{eq:rate_y_1}\\\nonumber
  R_{{s}} = \mathbb{E}[-{\log_2}p_{\hat{l}_s|\hat{z}_s,\hat{l}_b}({\hat{l}_s}|\hat{z}_s,\hat{l}_b)]
  + \mathbb{E}[-{\log_2}p_{\hat{z}_s}({\hat{z}_s})] + R_{{b}} 
\end{align}
With this prior, we can estimate the probability of $\hat{l}_s$ more accurately, thus improving the R-D performance.

\section{Experiment}
\subsection{Training Details}
COCO \cite{lin2014microsoft} dataset is used for training, and all images are randomly cropped into ${256 \times 256}$ patches during training. We adopt Adam \cite{kingma2014adam} optimizer with a batch size of 16 to train our model. Our network is optimized for 50 epochs with an initial learning rate of $1{e^{-4}}$, and the learning rate is reduced to $1{e^{-5}}$ for the last 15 epochs. 

We use $\lambda$ to adjust the minimum bitrate, and use the function $w(\cdot)$ to adjust the bitrate range covered by the model. During training, $k$ is randomly selected as an integer in [0,192] to change the size of the bitstream received by the decoder. Our models are optimized with two quality metrics, i.e. MSE (mean squared error) and MS-SSIM (multiscale structural similarity) \cite{wang2003multiscale}. For the MSE models, $\lambda$ is set to 0.002. For the MS-SSIM models, $\lambda$ is set to 7.0.

\subsection{Rate-distortion Performance}
\noindent\textbf{Compare with Scalable Coding Models.} We evaluate the rate-distortion  performance on the Kodak image set \cite{kodak} with 24 uncompressed $768\times512$ images (more results on the high-definition data set will be given in the Appendix). The rate is measured by bits per pixel (bpp), and the quality is measured by PSNR and MS-SSIM. First, we compare performance with other scalable coding models. For traditional methods SVC and HSVC, coarse-grained scalable coding mode performs better than fine-grained coding mode, so we compare with the former. For the learning-based image compression model \cite{toderici2015variable,zhang2019learned,guo2019deep,jia2019layered,mei2021learning}, we also compare coarse-grained methods, because no fine-grained methods have appeared before. As shown in Figure \ref{subfigure_1} and \ref{subfigure_2}, whether evaluated by PSNR or MS-SSIM, the rate-distortion performance of our model surpasses the existing traditional methods and learned methods by a big step. Furthermore, our model is also much more scalable than existing models. The bitstream of DeepFGS is nearly continuously adjustable, so that DeepFGS can well decode the bitstream which is truncated at any position. For the convenience of comparison, in this section, we decode $l_s$ with 8 channels as the interval to draw the RD curve of our model. In particular, for the enlarged part in Figure \ref{subfigure_1}, this interval is 1 channel.
Note that due to text limitations, the larger bitrate range and higher bitrate of DeepFGS are not shown in Figure \ref{fig:RMSE-100run}, but will be discussed in the Appendix.

\noindent\textbf{Compare with Simulcast Coding Models.} The goal of simulcast coding is similar to scalable coding, but simulcast mode encodes each layer independently, and the rates sent by the server are the sum of these layers. For example, in order to switch between three quality levels $(Q_1, Q_2, Q_3)$, the simulcast mode needs to encode three independent streams $(R_1, R_2, R_3)$. To achieve the same function as the three-layer scalable coding, the bitrate of the base layer is $R_1$, and the enhancement layers are $R_1 + R_2, R_1 + R_2 + R_3$. We test two well-known learning-based single-layer models \cite{minnen2018joint,cheng2020image} and two traditional single-layer models BPG \cite{bpgurl} and VVC \cite{VVC} in simulcast mode. As shown in figure \ref{subfigure_3}, our model performs best, and the advantage becomes more obvious as the number of layers increases. This is because there are lots of redundancies between the bitstreams of different layers in the simulcast mode.

\noindent\textbf{Compare with Non-Scalable Coding Models.} Due to the existence of redundancy between layers, compared to non-scalable coding, the performance degradation brought by the scalable method is inevitable. As shown in Figure \ref{subfigure_4}, we compared the performance with the non-scalable model (the same architecture as ours). It can be seen that the quality loss caused by our scalable method is very small. It is worth mentioning that our model is the first scalable model that exceeds BPG in PSNR.


\subsection{Ablation Study}
\label{ablation}
In order to evaluate the contribution of each module to the performance of our model, we perform ablation studies shown in Figure \ref{ablation_pic}. We study the following models: feature-level redundancy removal module (FRR), feature fusion module (FFM) and mutual entropy model (MEM). Table \ref{table1} shows the specific ablation configuration.The experimental results show that these modules all contribute to the performance improvement of DeepFGS, and combining them will achieve the best performance. In addition, by comparing the combinations $(Case2, Case3)$ and $(Case4, Case5)$, it can be found that MEM is more effective at high bitrates, while FFM is more effective at low bitrates.

\subsection{Visual Results}
As shown in Figure \ref{visual_results}, we present some cropped reconstructed images on the Kodak dataset. Compared with traditional \cite{wiegand2007text} and learning-based \cite{jia2019layered} scalable coding models, our model has two main advantages. Firstly, our method is able to use fewer bitrates to generate more visually pleasing results. For example, in contour and textural regions (e.g. human face and sweater), our method offers much more structural details than other codecs. Secondly, our method can generate reconstructed images of different qualities in fine-grained manner. As shown in the figure, our DeepFGS can provide dozens of continuous quality levels in the interval where \cite{wiegand2007text} and \cite{jia2019layered} can only generate two decoded images. More subjective quality results are given in the Appendix.

\subsection{Limitation}
The main limitation of our framework is that the information rearrangement strategy conflicts with the existing pixel-level autoregressive entropy model. The existing SOTA non-scalable image compression method utilizes this entropy model and trades decoding complexity for much rate-distortion performance. The study we are advancing is to design an autoregressive model compatible with the information rearrangement strategy. It is worth mentioning that our DeepFGS improves the performance of fine-grained scalable image compression from far inferior to BPG to beyond BPG. However, the performance will be upgraded to the VVC level once we complete the compatible autoregressive models.

\section{Conclusion}
This paper proposes the first learned fine-grained scalable coding model, DeepFGS, and achieves optimal compression performance and scalability. Ablation studies demonstrate the effectiveness of DeepFGS modules. In the following research, we will further explore the scalability of the image compression model, such as reducing the minimum tunable interval of the bitstream or increasing the rate range covered by the model.

\clearpage
{\small
\bibliographystyle{ieee_fullname}
\bibliography{reference.bib}
}
 
\clearpage
\section{Appendix}

\subsection{Introduction}
This Appendix contains many additional experiments and explorations, such as rate-distortion performance at high bitrates, models for wide bitrate ranges, and discussions on extreme fine-grained scalability. In addition, the appendix also includes the specific network structure of our DeepFGS model, as well as some high-definition image reconstruction results.


\subsection{Supplementary Note on Experiment}
All the experiments are conducted on the $Tesla\,V100$ GPU and $Intel\,Xeon\,Platinum\,8163$ CPU. Our framework DeepFGS are implemented using CompressAI \footnote{\url{https://github.com/InterDigitalInc/CompressAI}} PyTorch library. In addition, we used Pytorch version 1.9.1 and CompressAI version 1.1.7.

\subsection{Rate-Distortion Performance at High Bitrates}

\begin{figure}[h]
    \centering
  \begin{subfigure}[]{0.42\textwidth}
        \centering
      \includegraphics[width=1\textwidth]{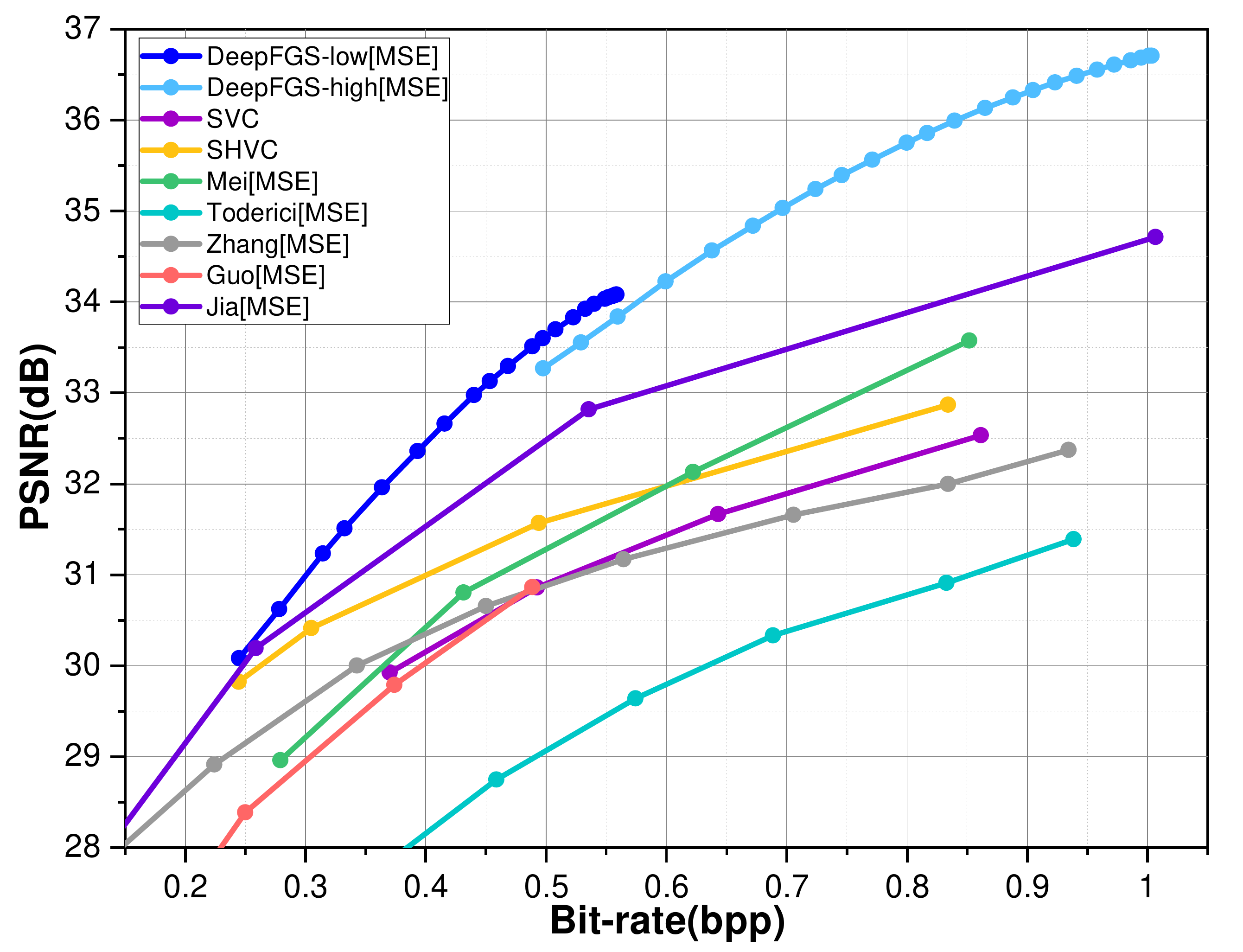} %
        \caption{}
        \label{PSNR-high}
    \end{subfigure}
  \begin{subfigure}[]{0.42\textwidth}
        \centering
      \includegraphics[width=1\textwidth]{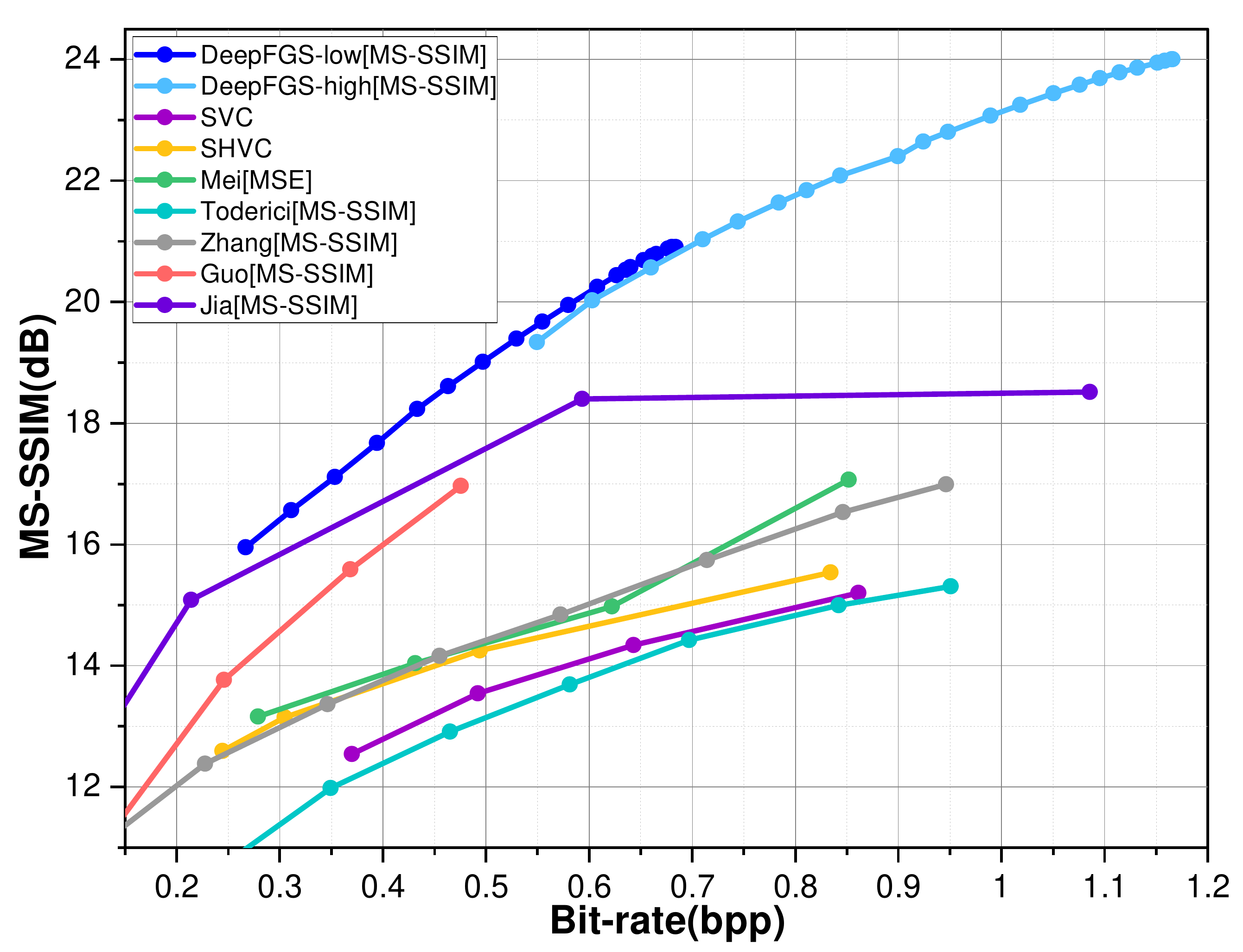}
        \caption{}
        \label{MSSSIM-high}
    \end{subfigure}
\caption{Comparison of rate-distortion performance of our DeepFGS with existing compression standards and other learned compression models at high bitrates, including Toderici \cite{toderici2015variable}, Zhang \cite{zhang2019learned}, Guo \cite{guo2019deep}, Jia \cite{jia2019layered}, Mei \cite{mei2021learning}. (a): PSNR performance evaluation on Kodak dataset. (b): MS-SSIM performance evaluation on Kodak dataset.}
\label{RD-high}   
\end{figure}

As shown in Figure \ref{RD-high}, we show the rate-distortion performance of our model at high bit rates. It can be observed that our DeepFGS model also has excellent performance at high bit rates. In addition, as the bit rate increases, the growth of PSNR has not slowed down, while the R-D curves of other scalable compression models tend to be flat.

\begin{figure}
  \centering
  \includegraphics[scale=0.26]{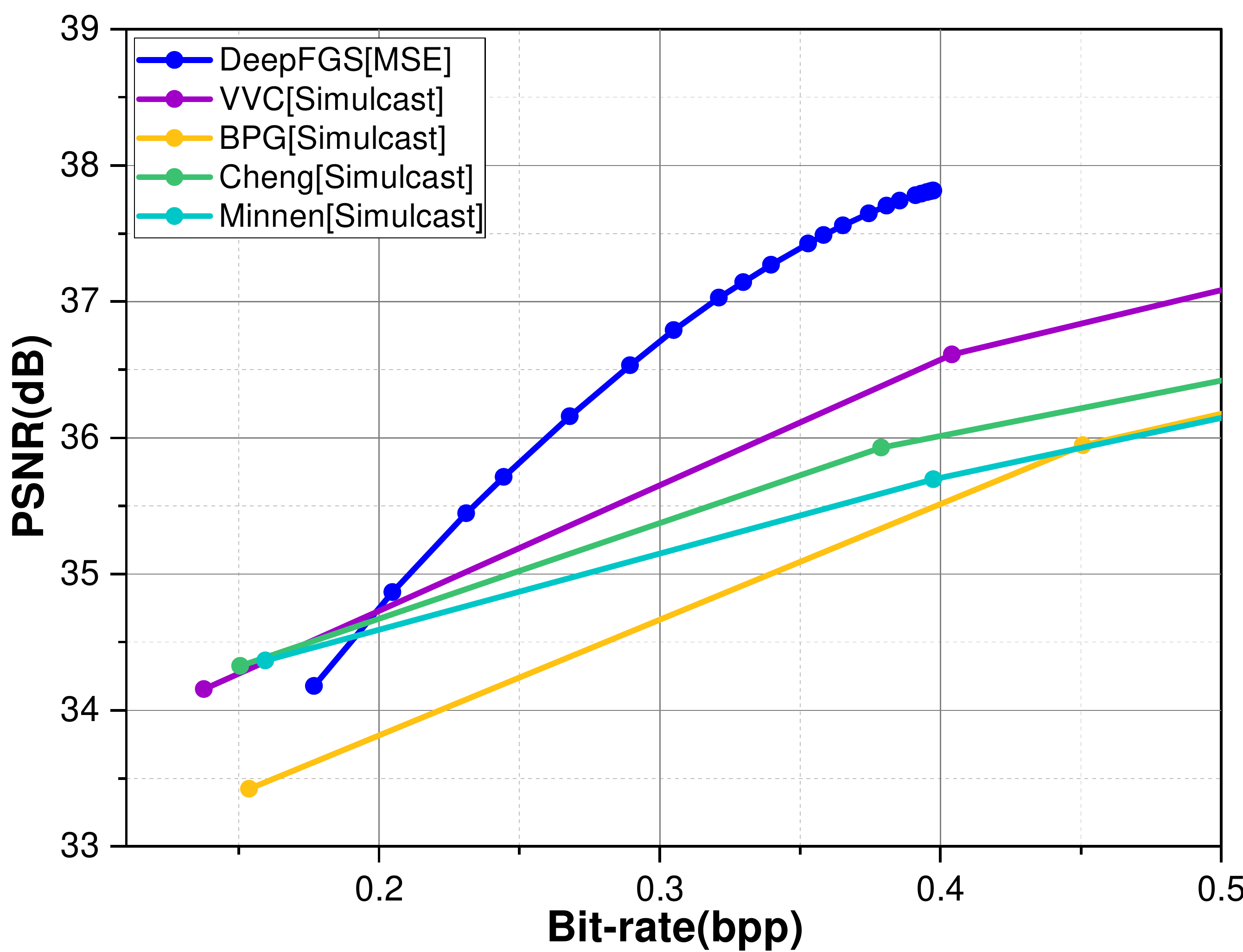}
  \caption{Comparison of rate-distortion performance at CLIC dataset.}
  \label{clic}
\end{figure}

We also evaluated the performance of DeepFGS on the high-definition dataset CLIC \footnote{\url{http://www.compression.cc}} Professional Validation dataset. In order to simplify the experiment, we directly show the broadcast performance of DeepFGS and other non-scalable models. As shown in Figure \ref{clic}, our model also performs well on the high-definition dataset.

\subsection{Discussion on Scalability}
\subsubsection{Discussion on the Scalable Range of Bitrate}
As shown in the experimental chapter, our DeepFGS-low and DeepFGS-high can provide scalability in the range of 0.5 bpp. The information rearrangement strategy constrains this scalability range:
\begin{align}
 &  J_{scalable} =R(l_b)+\lambda D(x,g_s(l_b)) + \\\nonumber 
 &  \qquad  R(l_b) +R(l_s^j) + \lambda w(j) D(x,g_s(l_b || l_s^j))\\\nonumber 
\label{appendix_eq1}
\end{align}

\begin{figure}
  \centering
  \includegraphics[scale=0.26]{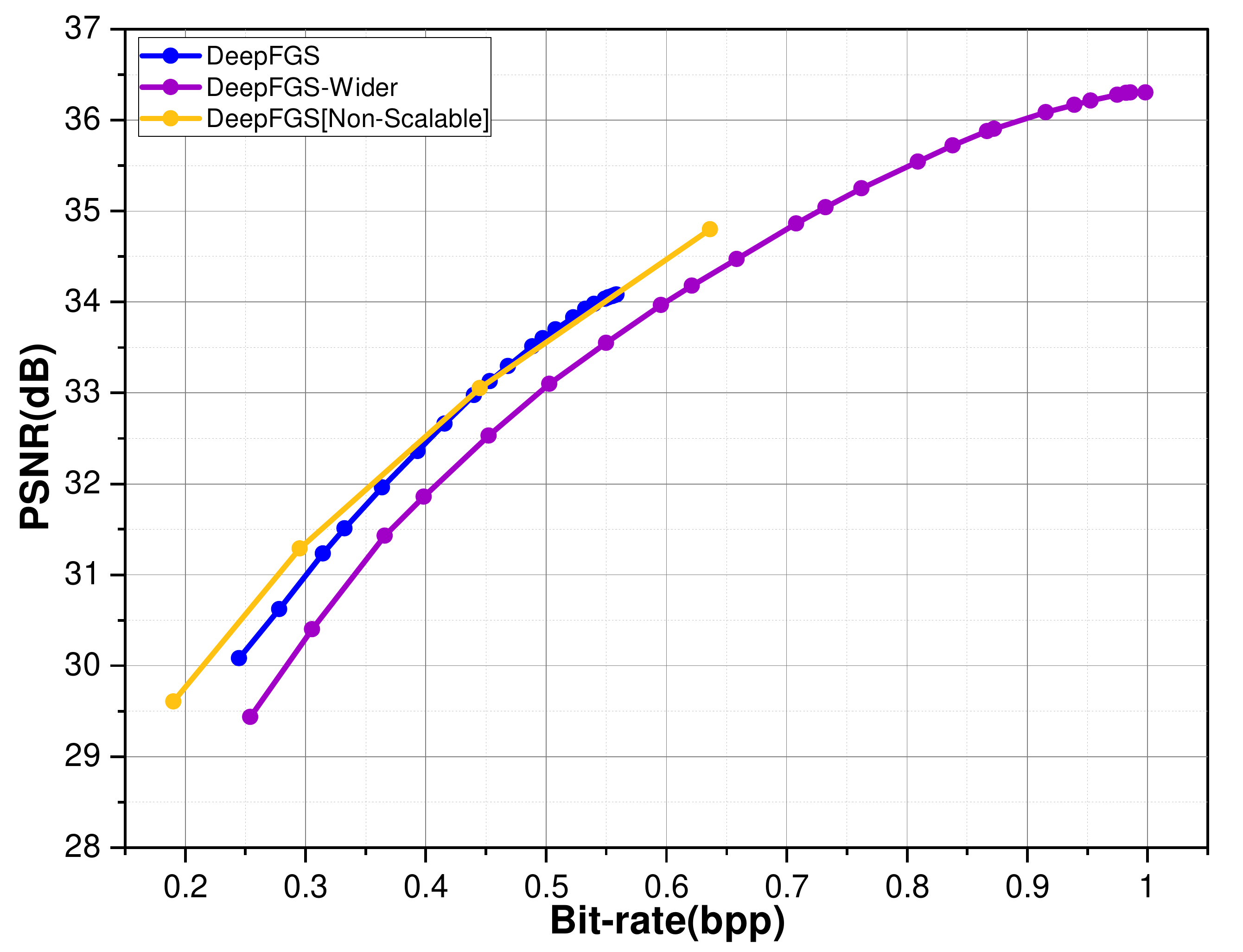}
  \caption{The impact of scalable ranges of bitrate on rate-distortion performance. Including the comparison between $w(j)=jMod8$ (DeepFGS) and $w(j)=j^2Mod64$ (DeepFGS-Wider).}
  \label{wider}
\end{figure}

Where $l_s^j$ refers to the first $j$ channels in $l_s$. $w(j)$ is a function that is positively correlated with $j$, so that when decoding more features (channels), the constraint on the bit rate can be relaxed. In this way, the decoder can learn to reconstruct reconstructed images of different qualities simultaneously. As shown in Figure \ref{wider}, we take two different functions of $w(j)=jMod8$, $w(j) = j^2Mod64$ as examples, and the scalable range of bitrate varies with $w(j)$. It can be observed that the wider the scalable range of the model, the more significant the quality loss (compared to the non-scalable model). Therefore, we chose $w(j)=jMod8$ with almost no loss in quality, and a scalable range of 0.5bpp is sufficient to cope with most bandwidth fluctuations.

\subsubsection{The Finest-grained Scalability}

\begin{figure}
  \centering
  \includegraphics[scale=0.26]{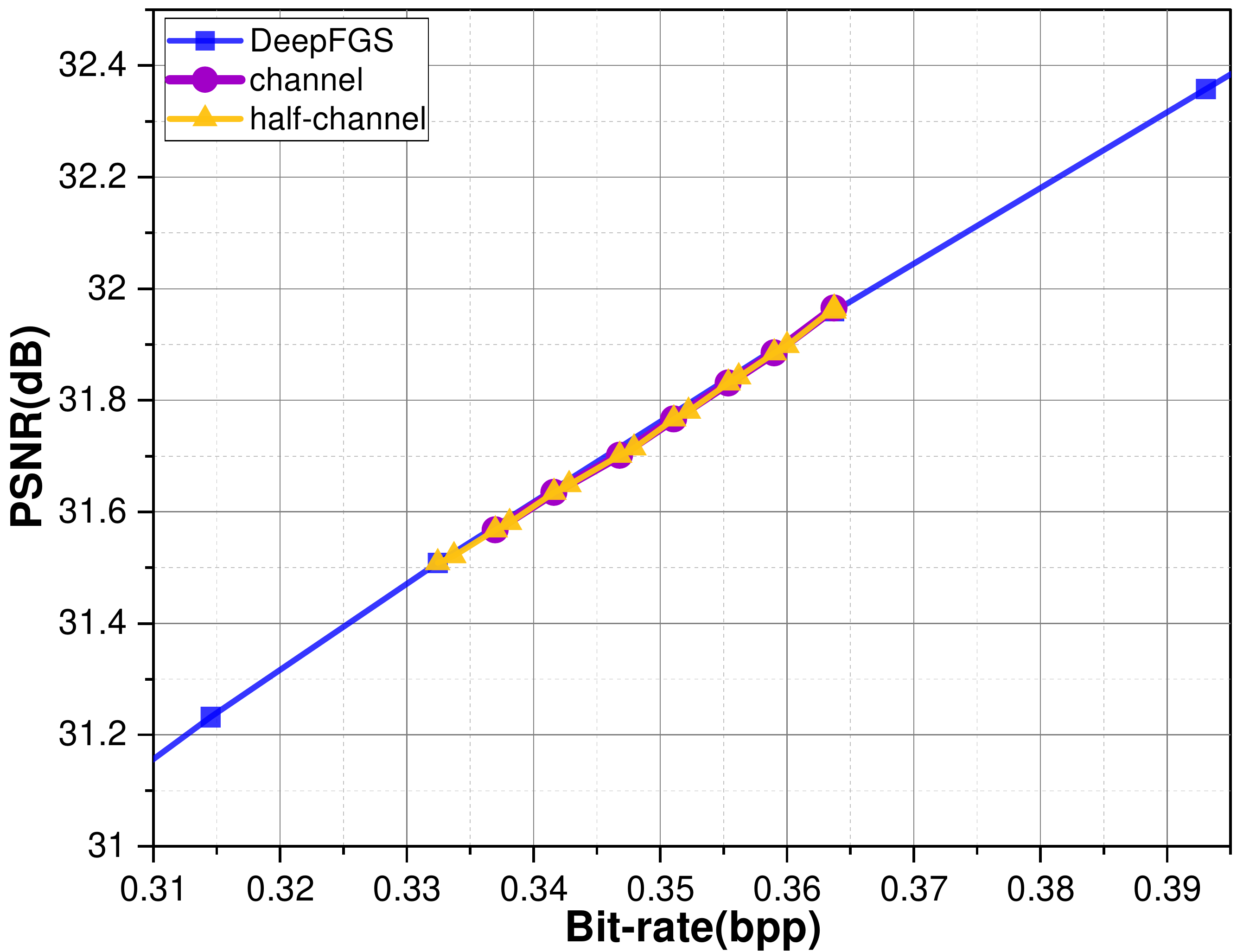}
  \caption{Comparison of channel-by-channel decoding and half-channel decoding.}
  \label{half}
\end{figure}

In the experimental section, we demonstrated the scalability of DeepFGS through channel-by-channel decoding. The same bitstream can reconstruct 192 images of different quality levels. Figure \ref{half} shows our further expansion, half-channel decoding (remaining channels with zero padding), which can achieve $192\times2$ quality levels of reconstruction. It can be observed that the performance of half-channel decoding is still continuous and stable. However, we can even get the finest scalability by decoding pixel by pixel, that is, $192\times\frac{w}{16}\times\frac{h}{16}$ quality levels.

In addition, we demonstrated the continuous quality improvement brought by channel-by-channel decoding in the supplementary material.

\begin{figure*}[h]
\begin{center}
\includegraphics[width=1.00\linewidth]{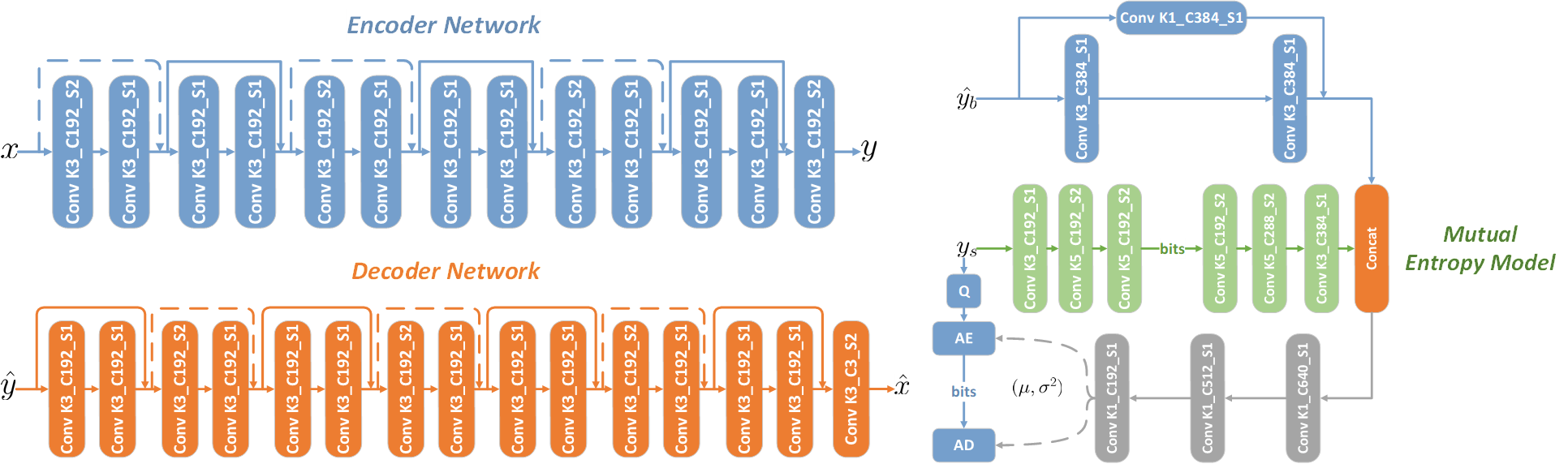}
\end{center}
  \caption{Architecture of network in DeepFGS, where Encoder Network refers to basic encoder or scalable encoder, and Decoder Network refers to  multiplexed decoder (excluding information fusion module). ”K” represents kernel size, ”C” denotes number of filters, and ”S” is stride of a convolutional layer. }
\label{network_detail}
\end{figure*}

\subsection{Network Details}
Figure \ref{network_detail} shows the details of our DeepFGS network structures. We use residual blocks and subpixel convolution to improve our backbone network. Specificly, we use 3x3 stacked convolutions with residual connection instead of 5x5 convolution, which can achieve larger receptive field while reducing the number of parameters. We use subpixel convolution in the upsampling layers. And after each upsampling or downsampling layer, generalized division normalization (GDN) layer is used as the activation function. In additional, The basic encoder and the scalable encoder use the same network structure.

\subsection{Additional Sample Image}
In the following pages, we will show more sample images from the Kadak dataset and CLIC Professional Validation dataset. In order to show the continuous improvement in the quality of our DeepFGS reconstructed images, we use different subsets of the same bitstream for decoding.

\begin{figure*}[h]
\begin{center}
\includegraphics[width=0.95\linewidth]{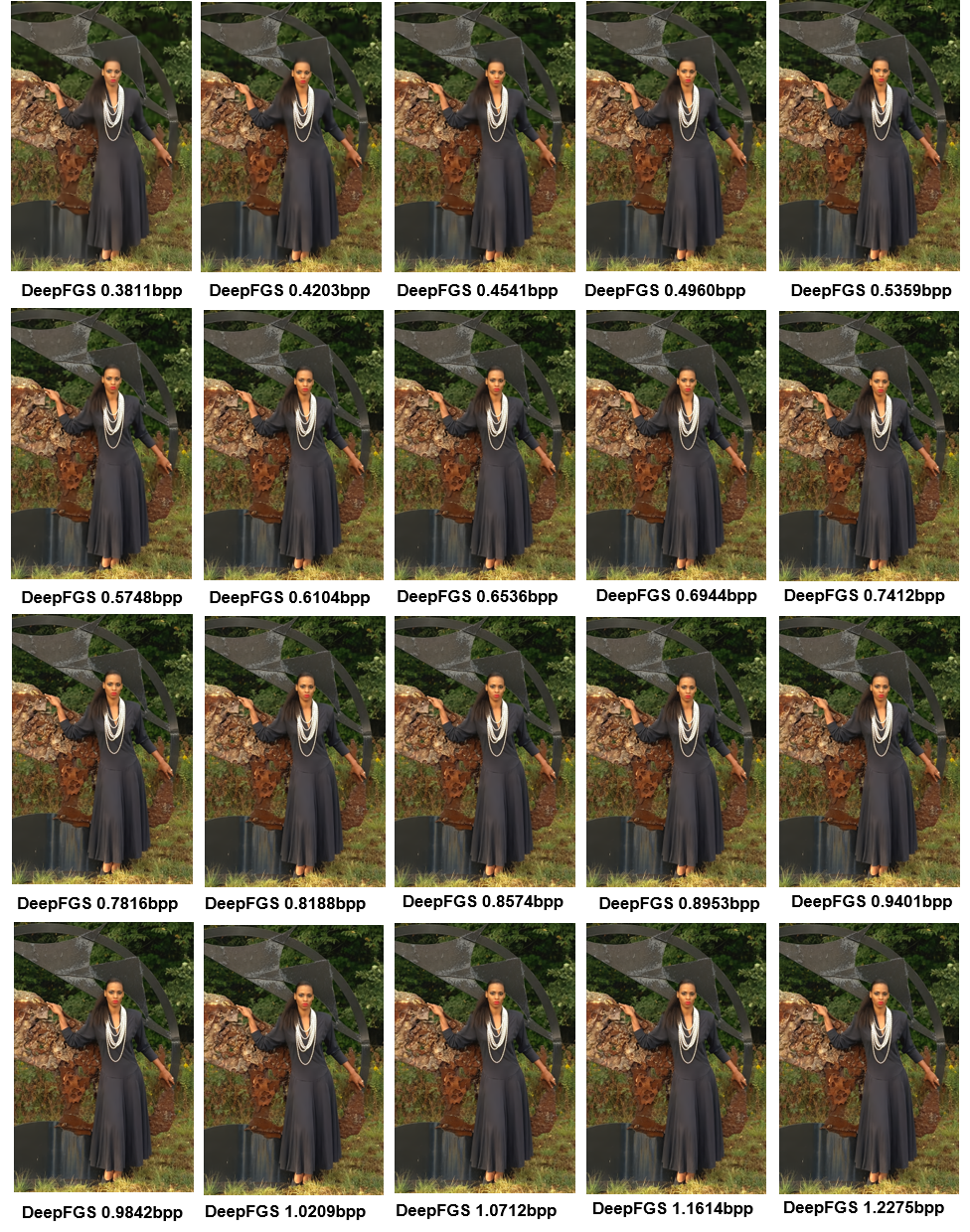}
\end{center}
  \caption{Visualization of reconstructed images \emph{kodim18} from Kodak dataset.}
\label{va1}
\end{figure*}

\begin{figure*}[h]
\begin{center}
\includegraphics[width=0.95\linewidth]{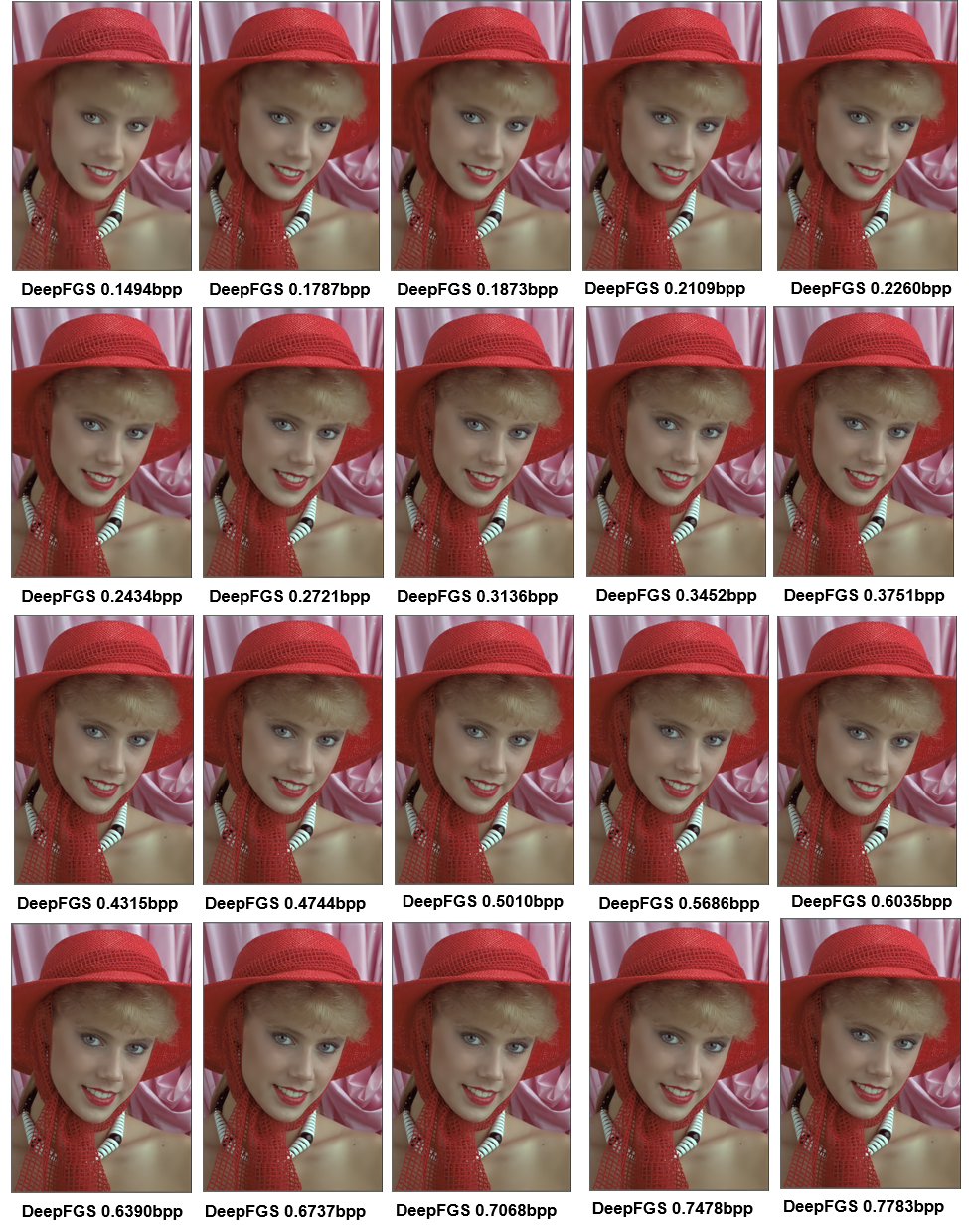}
\end{center}
  \caption{Visualization of reconstructed images \emph{kodim04} from Kodak dataset.}
\label{va1}
\end{figure*}

\begin{figure*}[h]
\begin{center}
\includegraphics[width=0.87\linewidth]{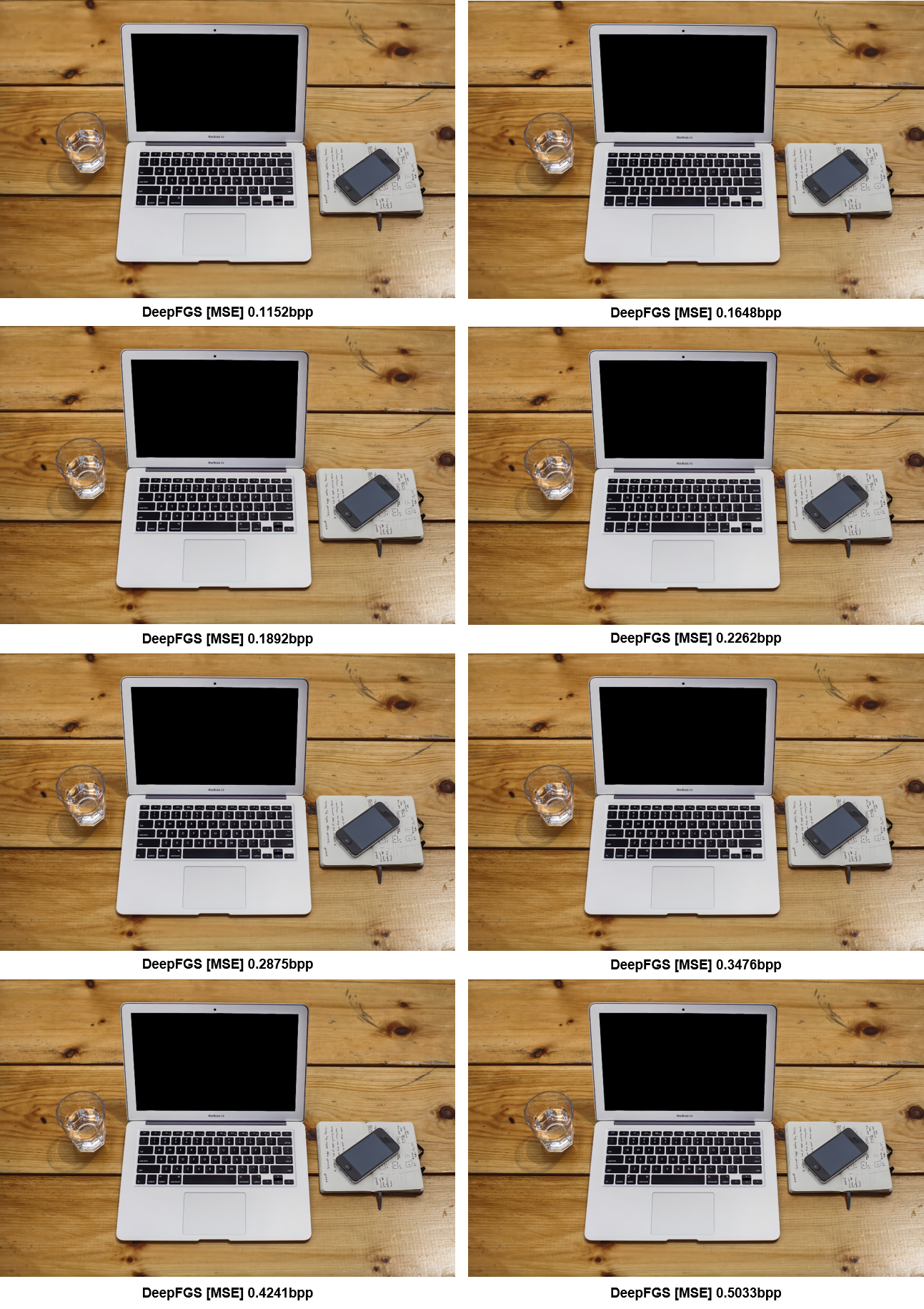}
\end{center}
  \caption{Visualization of reconstructed images \emph{alejandro-escamilla-6} from CLIC dataset.}
\label{va1}
\end{figure*}

\begin{figure*}[h]
\begin{center}
\includegraphics[width=0.87\linewidth]{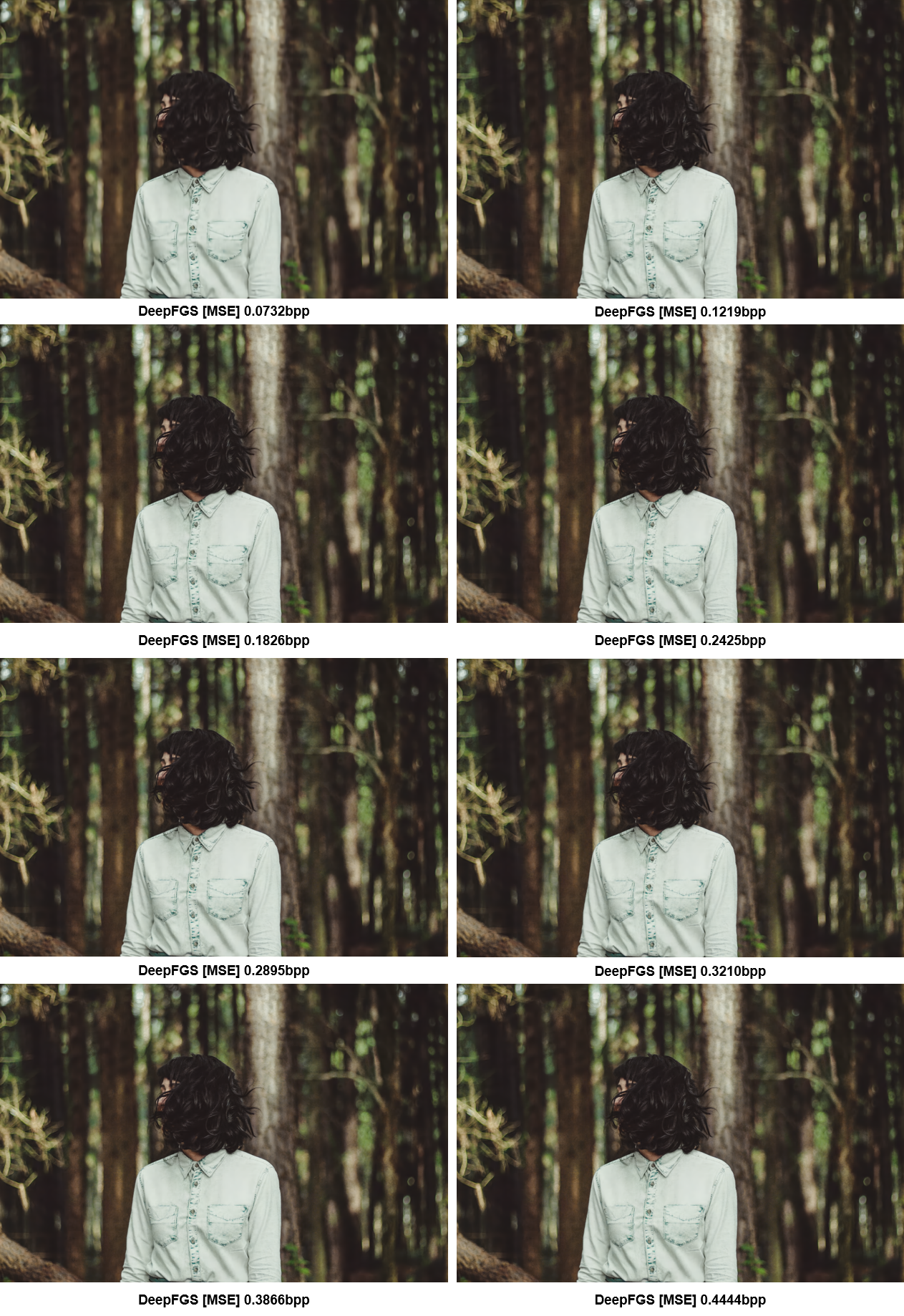}
\end{center}
  \caption{Visualization of reconstructed images \emph{allef-vinicius-109434} from CLIC dataset.}
\label{va1}
\end{figure*}

\begin{figure*}[h]
\begin{center}
\includegraphics[width=0.87\linewidth]{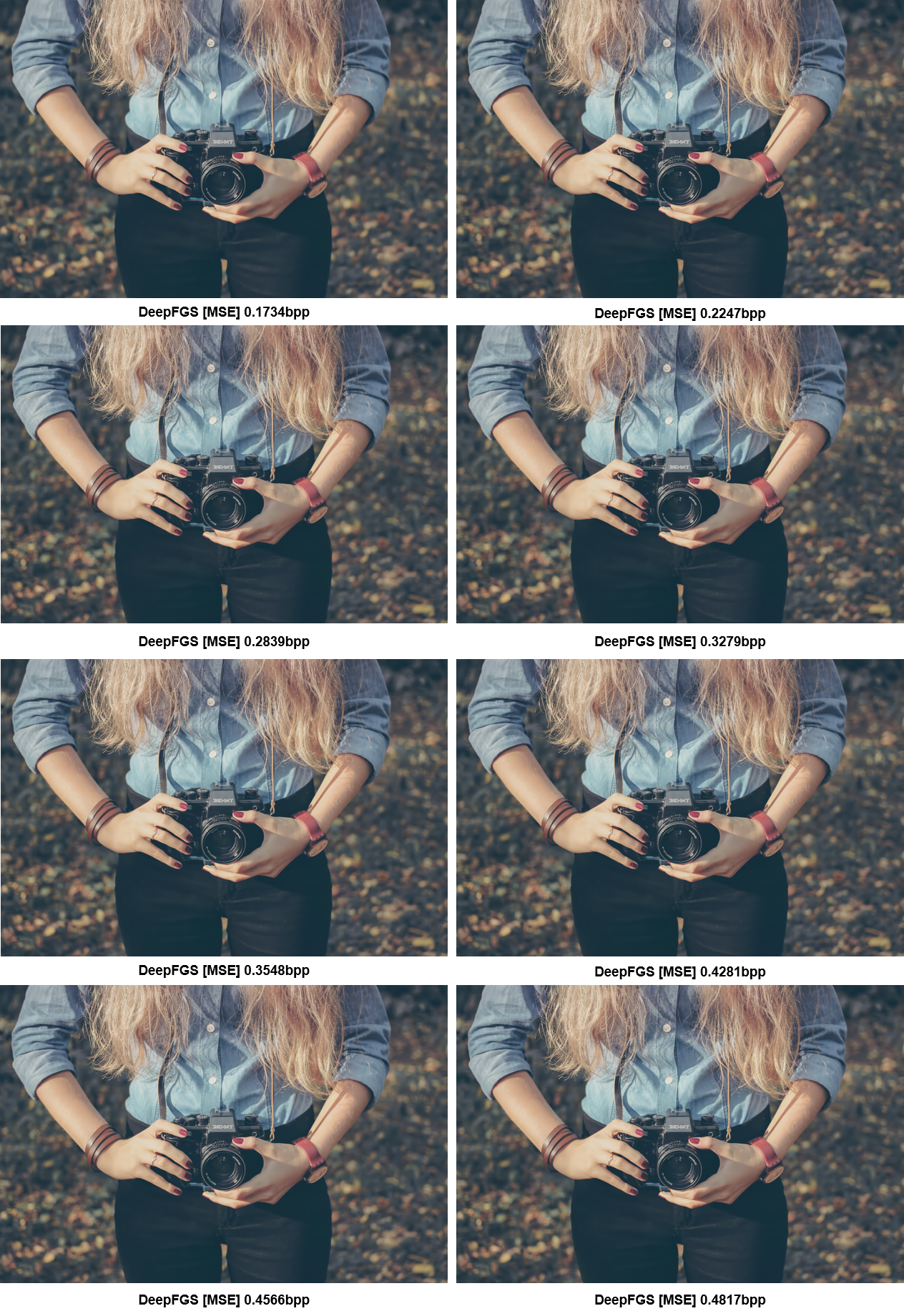}
\end{center}
  \caption{Visualization of reconstructed images \emph{sergey-zolkin-1045} from CLIC dataset.}
\label{va1}
\end{figure*}

\end{document}